\documentclass{aa}  
\usepackage{graphicx}
\usepackage{txfonts}
\begin{document}
   \title{The N$_2$D$^+$/N$_2$H$^+$ ratio as an evolutionary tracer of Class 0 protostars.\thanks{Based on observations with the IRAM~30m telescope.}}

   \subtitle{}

   \author{M. Emprechtinger
          \inst{1}
          \and P. Caselli \inst{2} \and N. H. Volgenau \inst{1}\fnmsep\thanks{Current address: California Institute of Technology, Owens Valley
Radio Observatory, Big Pine, CA 93513   USA} \and J. Stutzki \inst{1} \and M. C. Wiedner \inst{1}
          }

   \offprints{M. Emprechtinger}

   \institute{Universit\"at zu K\"oln,
              Z\"ulpicher Str. 77, 50937 Cologne\\
              \email{emprecht@ph1.uni-koeln.de}
	     \and
	     {School of Physics and Astronomy, University of Leeds, Leeds LS2 9JT, UK}
	      }

   \date{}

% \abstract{}{}{}{}{} 
% 5 {} token are mandatory
 
  \abstract
  % context heading (optional)
  % {} leave it empty if necessary  
   {Deuterated ions, especially H$_2$D$^+$ and N$_2$D$^+$, are abundant in cold ($\rm\sim 10~K$), dense ($\rm\sim 10^5~cm^{-3}$) regions, in which CO is frozen out onto  dust grains. In such environments, the N$_2$D$^+$/N$_2$H$^+$ ratio can exceed the elemental abundance ratio of D/H by a factor of $\simeq 10^4$.}
  % aims heading (mandatory)
   {We use the deuterium fractionation to investigate the evolutionary state  of Class 0 protostars. In particular, we expect the N$_2$D$^+$/N$_2$H$^+$ ratio to decrease as temperature (a sign of the evolution of the protostar) increases.}
  % methods heading (mandatory)
   {We observed N$_2$H$^+$~1-0, N$_2$D$^+$~1-0, 2-1 and 3-2, C$^{18}$O~1-0 and HCO$^+$~3-2 in a sample of 20 Class~0 and borderline Class~0/I protostars. We determined the deuteration fraction and searched for correlations between the N$_2$D$^+$/N$_2$H$^+$ ratio and well-established evolutionary tracers, such as T$\rm _{Dust}$ and the CO depletion factor. In addition, we compared the observational result with a chemical model.}
  % results heading (mandatory)
   {In our protostellar sample, the N$_2$H$^+$~1-0 optical depths are significantly lower than those found in prestellar cores, but the N$_2$H$^+$ column densities are comparable, which can be explained by the higher temperature and larger line width in protostellar cores. The deuterium fractionation of N$_2$H$^+$ in protostellar cores is also similar to that in prestellar cores. We found a clear correlation between the N$_2$D$^+$/N$_2$H$^+$ ratio and evolutionary tracers. As expected, the coolest, i.e. the youngest, objects show the largest deuterium fractionation. Furthermore, we find that sources with a high N$_2$D$^+$/N$_2$H$^+$ ratio show clear indication for infall (e.g. $\rm \delta v<0$). With decreasing deuterium fraction the infall signature disappears and $\rm \delta v$ tends to be positive for the most evolved objects. The deuterium fractionation of other molecules deviates clearly from that of N$_2$H$^+$. The DCO$^+$/HCO$^+$ ratio stays low at all evolutionary stages, whereas the NH$_2$D/NH$_3$ ratio is $>0.15$ even in the most evolved objects.}
  % conclusions heading (optional), leave it empty if necessary 
   {The N$_2$D$^+$/N$_2$H$^+$ ratio is known to trace the evolution of prestellar cores; we show that this ratio can be used to trace core evolution even after star formation. Protostars with an N$_2$D$^+$/N$_2$H$^+$ ratio above 0.15 are in a stage shortly after the beginning of collapse. Later on, deuterium fractionation decreases until it reaches a value of $\sim 0.03$ at the Class~0/I borderline.}


   \maketitle
%
%________________________________________________________________

\section{Introduction}\label{intr}

Early attempts to describe an evolutionary sequence of protostars classified them according to their near- and mid-infrared spectra.  Lada \& Wilking~(\cite{class}) divided a sample of protostellar sources in the $\rho$~Ophiuchi cloud into three classes (Class I to III with progressive evolutionary stage). Andr\' e et al.~(\cite{Class0}) introduced the Class~0 stage. The strong submillimetre emission of Class~0 objects relative to their bolometric luminosity suggests that the mass of the circumstellar envelope is larger than the mass of the central star. Class~0 objects are the youngest among the protostars. In the subsequent years, more sensitive indicators for protostar evolution have been established, such as the bolometric temperture (Myers \& Ladd~\cite{tbol}), the L$\rm _{BOL}$/L$\rm _{smm}$ ratio (Andr\' e et al.~\cite{Class0}) and the correlation between the bolometric luminosity and the distance normalized 1.3~mm flux (Saraceno et al.~\cite{sar}). All these evolutionary indicators are based on the change of the spectral energy distribution (SED) of the protostar. As it heats up, an increasing fraction of the radiation is emitted at shorter wavelengths, leading to an increase of L$\rm _{BOL}$ and T$\rm _{BOL}$ with time.

Several theoretical models connect physical properties of the protostars with the time since gravitational collapse (e.g. Smith~\cite{mod1}, Myers et al.~\cite{mod2}). However, although the evolutionary stages determined by different models is the same (i.e. the time sequence is the same), the absolute ages vary significantly.  The absolute age for Class~0/I borderline objects, for example, varies between $10^4$  and a few times $10^5$  years (Froebrich~\cite{fro}). In this paper, we show that it is possible to use the deuterium fractionation of N$_2$H$^+$ as an evolutionary tracer.

The chemistry of cold ($\rm T<20~K$), dense ($\rm n=10^5~cm^{-3}$) environments in space is peculiar. Common molecular tracers, such as CO and CS, appear reduced in abundance because they freeze out onto dust grains (e.g. Willacy et al.~\cite{wlv98}, Caselli et al.~\cite{cwt99}, Kramer et al.~\cite{kal99}, Bergin et al.~\cite{ted}, Redman et al.~\cite{rrn02}, Tafalla et al.~\cite{taf},~\cite{tmc04}). By comparison, molecules such as N$_2$H$^+$ and NH$_3$ suffer less from depletion. The persistence of nitrogen-bearing species is likely due to the fact that N$_2$, the progenitor of these species, remains in the gas phase longer than CO, despite the fact that recent laboratory measurements have found that the binding energy, as well as the sticking coefficient of N$_2$ and CO on dust, are quite similar ({\"O}berg et al.~\cite{ovf05}, Bisschop et al.~\cite{bis}).  Recent theoretical work suggests that the resistance of nitrogen-bearing species to depletion may be due to the fact,
  that a significant fraction of interstellar nitrogen is in atomic (N) rather than molecular (N$_2$) form (Maret et al.~\cite{mar}, Flower et al.~\cite{flo}). Atomic nitrogen has a smaller binding energy than molecular nitrogen and CO.

One consequence of the low temperatures and freeze-out of CO in pre- and protostellar cores is that the abundance of deuterated molecules is greatly enhanced. In such regions, column density ratios for N$_2$D$^+$/N$_2$H$^+$ of 0.24 (Caselli et al.~\cite{cas2}), D$_2$CO/H$_2$CO $\simeq$ 0.01-0.1 (Bacmann et al. \cite{bac03}), NH$_2$D/NH$_3$ up to 0.33 (Hatchell~\cite{nh3})  and DCO$^+$/HCO$^+$ in the order of 0.01 (J\o rgensen et al.~\cite{jor}) are found. These ratios are three to four orders of magnitudes larger than the average ISM D/H ratio of 1.5$\cdot10^{-5}$ (Oliveira et al.~\cite{dh}). Models show that the abundance of H$_2$D$^+$, a fundamental molecule in deuterium chemistry, rises steeply as temperature decreases (e.g. Flower et al.~\cite{fpfw}, Roberts \& Millar~\cite{rob}). Consequently, the deuterium fractionation of many other molecules also increases (Millar et al.~\cite{tom}). 

In a recent study, Crapsi et al.~(\cite{crap}) investigated the N$_2$D$^+$/N$_2$H$^+$ ratio in 31 low-mass $starless~cores$ and found ratios in the range of a few percent to 0.44. They also found a tight correlation between the N$_2$D$^+$/N$_2$H$^+$ ratio and the CO depletion factor; the ratio is greater where CO is depleted. Furthermore, they attempted to find correlations between the N$_2$D$^+$/N$_2$H$^+$ ratio and various chemical and dynamical evolutionary indicators (e.g. core density, molecular column density, line width, and line asymmetry). Although the dependencies on particular indicators are not always clear, they identified a recognizable trend. The N$_2$D$^+$/N$_2$H$^+$ ratio was generally found to increase as a cloud core evolves towards the moment of protostellar collapse. Another results from their work is that the trends seen in the subsample of Taurus cores were more homogeneous than trends in the total sample. This homogeneity led them to speculate that a core's external environment plays an important role in its evolution. In this work we follow the trend of the N$_2$D$^+$/N$_2$H$^+$ ratio to the subsequent, early protostellar stages.

One recent study that also sought to determine the behavior of deuterated molecules in protostellar environments is from Roberts \& Millar~(\cite{RM}). They determined the N$_2$D$^+$/N$_2$H$^+$ ratio in five low mass $protostellar~cores$, using the University of Arizona 12m telescope (HPBW of 70$''$ for N$_2$H$^+$~1-0). Furthermore they measured the D$_2$CO/H$_2$CO and the HDCO/H$_2$CO ratio in these cores. The HDCO/H$_2$CO ratio was similar in most cores ($\sim 0.06$), with the exception of the object with the highest N$_2$D$^+$/N$_2$H$^+$ ratio,  where HDCO/H$_2$CO was lowest. For four sources, the N$_2$D$^+$/N$_2$H$^+$ ratio appeared anti-correlated with the D$_2$CO/H$_2$CO ratio. Only in L~1527 were both N$_2$D$^+$ and D$_2$CO abundances anomalously low. They did not find a correlation between deuterium fractionation and bolometric temperature for any of the observed molecules. However, due to the small number of protostars, this work cannot rule out any trend of the deuterium fractionation in protostellar cores. 

We observed 20 Class~0 protostars in N$_2$H$^+$ and N$_2$D$^+$, using the IRAM~30m telescope (HPBW of 27$''$ for N$_2$H$^+$~1-0, see section~\ref{obs}).
The goal of our study is to investigate the trend of deuterium fractionation after the moment of star formation. Spectra and line parameters are presented in Sect.~\ref{res}. We expect that the N$_2$D$^+$/N$_2$H$^+$ ratio (determined in section~\ref{ana1}) should decrease during protostellar evolution due to the internal heating of the core. As the core environment warms up, H$_2$D$^+$ and its more highly deuterated isotopologues are destroyed.  Consequently, all other deuterated molecules formed by reactions with H$_2$D$^+$ (such as N$_2$D$^+$) are destroyed as well. In Sect.~\ref{Tdustsec}-~\ref{ana4}, the N$_2$D$^+$/N$_2$H$^+$ ratio is compared with evolutionary indicators, namely dust temperature, CO depletion factor, and L$\rm _{BOL}$/F$_{1.3}$. Possible correlations with kinematical parameters of the gas are investigated in section~\ref{anav}. In addition, we compare the N$_2$D$^+$/N$_2$H$^+$ ratio with the deuterium fractionation of other molecules (section~\ref{anam}). Finally, we compare our results with the predictions of a simple chemical model (section~\ref{model}). 

\section{Observations}\label{obs}

\subsection{Source sample}

Our survey includes 20 Class~0 protostellar cores. Tab.~\ref{obj} lists the objects, their coordinates and their estimated distances. The sources were selected in terms of being a.) well known (i.e. many complementary data can be found in the literature) and b.) representing a range of stages of Class 0  sources, from very young objects (e.g. HH~211, which is one of the youngest protostars according to Froebrich~\cite{fro}) to borderline objects (Class 0/1), such as L1455~A1 (Froebrich~\cite{fro}) and Barnard~5 IRS~1 (Velusamy \& Langer~\cite{B5}).
For one object, L~1527, the classification is ambiguous. The continuum data denote it as a Class~0 object, but an embedded source is detected at a wavelength $< 5~\mu m$ (Froebrich~\cite{fro}), indicating that this protostar is already more evolved. Therefore, this source is also classified as a Class~0/I borderline object.
Most of the objects (13) are located in the Perseus molecular cloud, and all are at a distance closer than 500 pc. For the sources located in Perseus, we assumed a distances of 220~pc (\v{C}ernis~\cite{su}). Other authors (e.g. Herbig \& Jones~\cite{Per350}) give distances larger than 300~pc. Eleven objects belong to a sample of sources catalogued by Froebrich~(\cite{fro}), and for 17 sources, kinetic gas temperatures are known from NH$_3$ observations (Jijina et al.~\cite{jij}). However, the NH$_3$ observations vary in spatial resolution between 40$''$ and 80$''$. To avoid systematic errors, introduced by the different beam sizes, we use these 
kinetic temperatures here only as an estimate of the gas temperature.

\begin{table}[ht]
\caption{Class~0 objects observed for the survey.} \label{obj}
\begin{center}
\begin{tabular}{lccc}
\hline
Source & RA & DEC & Distance\\
& (J2000.0) & (J2000.0) & [pc] \\ 
\hline
\hline
L 1448 IRS 2 & 03:25:22.4 & 30:45:12 & 220$^1$\\
L 1448 IRS 3 & 03:25:36.4 & 30:45:20 & 220$^1$\\
L 1448 C & 03:25:38.8 & 30:44:05 & 220$^1$\\
L 1455 A1 & 03:27:42.1 & 30:12:43 & 220$^1$\\
L 1455 A2 & 03:27:49.8 & 30:11:42 & 220$^1$\\
Per 4 S & 03:29:17.9 & 31:27:30 & 220$^1$\\
Per 4 N & 03:29:22.7 & 31:33:30 & 220$^1$\\
Per 5 & 03:29:51.6 & 31:39:03 & 220$^1$\\
Per 6 & 03:30:14.8 & 30:23:48 & 220$^1$\\
IRAS 03282+3035 & 03:31:21.0 & 30:45:30 & 220$^1$\\
Barnard 1-b & 03:33:21.3 & 31:07:35 & 220$^1$\\
HH 211 & 03:43:56.8 & 32:00:50 & 220$^1$\\
Barnard 5 IRS1 &  03:47:41.6 & 32:51:42 & 220$^1$\\
L 1527 & 04:39:53.5 & 26:03:05 & 140$^2$\\
L 483 & 18:17:29.8 & -04:39:38 & 200$^2$\\
SMM 5 & 18:29:51.2 & 01:16:40 & 310$^3$\\
L 723 & 19:17:53.1 & 19:12:16 & 300$^2$\\
L 673 A & 19:20:25.8 & 11:19:51 & 300$^2$\\
B 335 & 19:37:01.4 & 07:34:06 & 250$^2$\\
L 1157 & 20:39:06.5 & 68:02:13 & 440$^4$\\
\hline
\end{tabular}
\end{center}
$^1$\v{C}ernis~(\cite{su}), $^2$Jijina et al.~(\cite{jij}), $^3$Froebrich~(\cite{fro}), $^4$Gueth et al.~(\cite{gue}).
\end{table}

\subsection{Molecular line observations}

The observations were carried out in January 2006 using the IRAM-30m telescope. In Tab.~\ref{lines}, the molecular line transitions, their frequencies, and their corresponding half power beam widths (HPBW) are given.
Two different receiver set-ups were used. The first set-up used frequency switching to observe C$^{18}$O~1-0, N$_2$D$^+$~2-1, N$_2$D$^+$~3-2 and HCO$^+$~3-2 simultaneously. The offsets were 10 (C$^{18}$O 1-0), 15 (N$_2$D$^+$~2-1), 25 (N$_2$D$^+$~3-2) and 28 (HCO$^+$~3-2) MHz, respectively. 
The second set-up used an external local oscillator to observe the N$_2$D$^+$~1-0 line. In this mode, frequency switching was not available, so position switching was used instead. The N$_2$D$^+$~1-0 line was observed simultaneously with the N$_2$H$^+$~1-0 and N$_2$D$^+$~3-2 lines. As a back end, the VESPA autocorrelator was used, which was adjusted to a resolution of about 0.05~km/s at each frequency.

\begin{table}[ht]
\caption{Observed emission lines.} \label{lines}
\begin{center}
\begin{tabular}{lcc}
\hline
Line & Frequency [GHz]& HPBW [$''$]  \\ 
\hline
\hline
N$_2$D$^+$ 1-0 & 77.109 & 32 \\
N$_2$H$^+$ 1-0 & 93.173 & 27 \\
C$^{18}$O 1-0 & 109.782 & 22 \\
N$_2$D$^+$ 2-1 & 154.217 & 16 \\
N$_2$D$^+$ 3-2 & 231.322 & 11\\
HCO$^+$ 3-2 & 267.558 & 10 \\
\hline
\end{tabular}
\end{center}
\end{table} 

During observations, the atmospheric $\tau$(225~GHz) was about 0.25, which is a typical opacity for winter weather conditions at the IRAM site. The corresponding system temperatures were between 150~K (N$_2$H$^+$~1-0) and 800~K (HCO$^+$~3-2).      
Pointing was checked every two hours, and the pointing accuracy was better than 3$''$. Antenna temperatures and main beam temperatures are related by the main beam efficiency, $\rm\eta_{MB}$, which is between  0.8 and 0.42, depending on the observing frequency\footnote{http://www.iram.fr/}.

\section{Results}\label{res}

\subsection{N$_2$H$^+$ \& N$_2$D$^+$}

\begin{figure}[ht]
   {\resizebox{8cm}{!}
            {\includegraphics{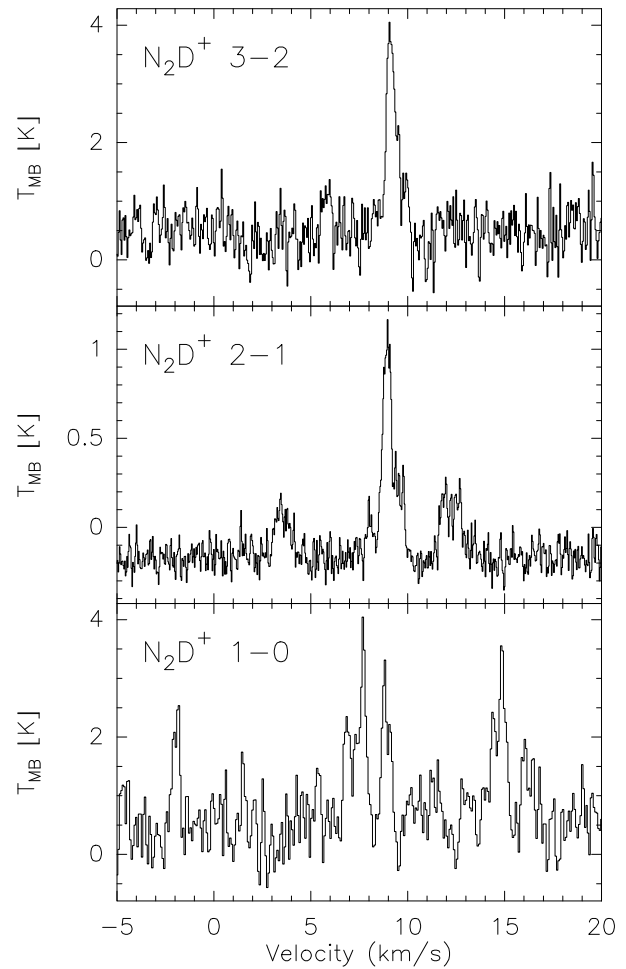}}}    
     \caption{Spectra of N$_2$D$^+$~1-0, N$_2$D$^+$~2-1 and N$_2$D$^+$~3-2 toward HH~211, one of the 
youngest sources in our sample.} \label{nspec}
      \end{figure}

\begin{figure*}[ht]
\begin{center}
   {\resizebox{13.5cm}{!}
            {\includegraphics{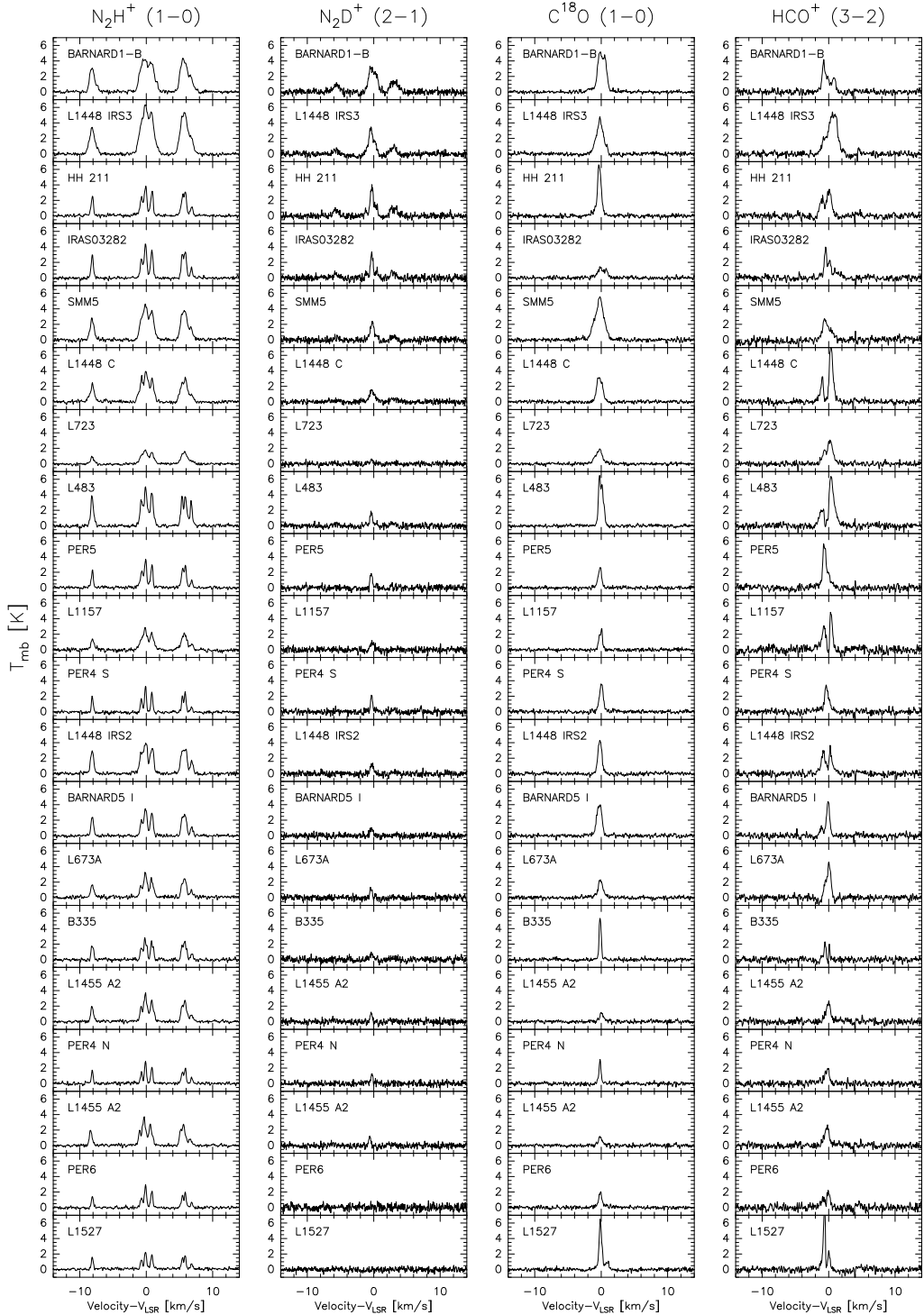}}}    
     \caption{Spectra of N$_2$H$^+$~1-0, N$_2$D$^+$~2-1, C$^{18}$O~1-0 and HCO$^+$~3-2 (from left to right) of all 20 protostars of our sample. To be able to plot all spectra in one plot we divided the intensities of HCO$^+$ by three.} \label{spec}
     \end{center}
      \end{figure*}

The results of the N$_2$H$^+$ and N$_2$D$^+$ observations are summarized in Tab.~\ref{n2h}. 
Fig.~\ref{nspec} shows the three N$_2$D$^+$ transitions observed toward HH~211, one of the few sources (total of 6, see Table~\ref{n2h}) where all lines are detected.  
The N$_2$H$^+$~1-0, N$_2$D$^+$~2-1, C$^{18}$O~1-0 and HCO$^+$~3-2 spectra of all 20 sources are displayed in Fig.~\ref{spec}.

\begin{table*}[ht]
\caption{Line parameters of the N$_2$H$^+$ and N$_2$D$^+$ lines.}\label{n2h}
\begin{center}
\begin{tabular}{lccccccccccc}
\hline
& \multicolumn{4}{c}{N$_2$H$^+$~(1-0)} &  &\multicolumn{6}{c}{N$_2$D$^+$} \\
\cline{2-5} \cline{7-12} 
Source & $W$ & $\Delta$v  & $\tau$ & T$\rm _{ex}$ & T$\rm _{Kin}$ & $W_{J=1-0}$& $W_{J=2-1}$  & $W_{J=3-2}$ & $\tau$ & $\rm\Delta v_{2-1}$ &  T$\rm _{ex}$ \\ 
 & [Kkm/s] & [km/s] & & [K] & [K] & [Kkm/s] & [Kkm/s] & [Kkm/s]& J=2-1 & [km/s] &[K] \\
\hline
\hline
L 1448 IRS 2 & 12.93 & 0.50 & 11.6 & 6.8 & - & 1.7 & 1.10 & 0.73 & $<$ 0.1 &    0.53 &                       -          \\
L 1448 IRS 3 & 24.65 & 0.93& 4.1 & 11.4 & 12.0 & 4.0 & 5.05  & 3.22 & 2.6 &   0.59 &                    7.9       \\
L 1448 C & 12.48 & 0.76& 5.3 & 7.3 & 13.7$^*$ & $<$ 1 & 2.32 & 1.38 & $<$ 0.1 &      0.72 &                       -          \\
L 1455 A1 & 7.20 &0.74 & 1.4 & 9.2 & 15.0 & $<$ 1 & 0.68 & $<$ 0.74 & $<$ 0.1 &  0.42 &               -     \\
L 1455 A2 & 8.60 &0.54 & 3.0 & 9.1 & 11.0 & $<$ 1 & 1.03 & $<$ 0.73 &  $<$ 0.1 & 0.33 &              -    \\
Per 4 S & 5.08 &0.29 & 3.5 & 8.5 & 9.0 & 1.3 &1.43 &  0.98 & $<$ 0.1 &     0.32 &                 -          \\
Per 4 N & 4.62 & 0.31& 3.6 & 7.6 & 10.0 &  1.3 & 0.75 & $<$ 0.68 &  $<$ 0.1 &    0.31 &           -       \\
Per 5 & 7.53 & 0.38 & 3.8 & 8.7 & 11.0 &$<$ 1 & 1.07  & 1.08 &  $<$ 0.1 &   0.36 &                 -            \\
Per 6 & 4.64 & 0.36 & $<$ 0.1 & - & 11.0 &$<$ 1 & $<$ 0.18 & $<$ 0.69 & - &  -   &                   -           \\
IRAS 03282 & 9.53 & 0.38& 5.3 & 8.6 & 11.3$^*$ & $<$ 1 & 3.58 & 1.47 & 3.9 &  0.34 &                    7.9         \\
Barnard 1-B & 19.05 &0.82 & 8.6 & 7.6 & 12.0 & 5.9& 7.48 & 4.82 & 4.9 &  0.78    &                   6.7          \\
HH 211 & 9.18 & 0.41& 6.2 & 7.6 & 12.8$^*$ & 2.3 & 5.24 & 2.08 & 4.9 &   0.39   &                    7.5              \\
Barnard 5 IRS 1 & 8.59 &0.43 & 6.4 & 7.2 & 10.8$^*$ & $<$ 1 & 0.50 & 0.65 & $<$ 0.1 & 0.50 &            -  \\
L 1527 & 3.91 & 0.32& 5.6 & 5.7 & 10.8$^*$ & $<$ 1& $<$ 0.18 & $<$ 0.59 & -  &   - &                    -           \\
L 483 & 14.05 & 0.44& 12.1 & 7.3 & 10.0 & 2.1 & 1.75  & 1.13 & 2.8 &     0.37   &                5.5              \\
SMM 5 & 15.39 &0.70 & 4.9 & 8.9 & 12.0 & $<$ 1 & 2.46 & 1.43 &  $<$ 0.1  &     0.58 &                       -           \\
L 723 & 5.70 & 0.71& 2.4 & 6.6 & 11.0 & $<$ 1 & 0.34 & 1.29 & $<$ 0.1  &     0.32 &                        -            \\
L 673 A & 7.34 &0.54 & 2.9 & 8.6 & $<$12.0 & $<$ 1 & 0.45 & 0.63 & $<$ 0.1  &    0.34 &                          -          \\
B 335 & 6.05 & 0.40& 9.0 & 5.7 & 9.9 & $<$ 1 & 0.42 & 0.42 & $<$ 0.1 &     0.46    &                        -               \\
L 1157 & 8.38 &0.65 & 2.1 & 8.9 & - & $<$ 1& 0.53 & 1.00 & $<$ 0.1 &      0.56 &                        -                \\
\hline
\end{tabular}
\end{center}
The kinetic gas temperatures marked with an $^*$ are taken from Hatchell~(\cite{nh3}). All other T$\rm _{kin}$ values are listed in Jijina et al.~(\cite{jij}). The errors on the integrated intensities of N$_2$H$^+$~1-0 and N$_2$D$^+$~2-1 are $\sim 10\%$ (calibration uncertainties), whereas the errors on the N$_2$D$^+$~1-0 and N$_2$D$^+$~3-2 lines are dominated by the noise of the spectra (0.5 and 0.2 Kkm/s for J=1-0 and J=3-2, respectively). $\tau$ is the total optical depth, defined as the sum of the optical depths of the individual hyperfine components. Typical errors of $\tau$ are $\pm 0.5$ and $\pm 0.2$ for N$_2$H$^+$ and N$_2$D$^+$~2-1 (where it is detected), respectively. The resulting error of T$\rm _{ex}$ is $\sim 0.2$~K for both species.
\end{table*}

The N$_2$H$^+$~1-0 line was detected in all sources of our sample. The total optical depth was determined by fitting all hyperfine components of the line using the "METHOD HFS" of the CLASS program\footnote{http://www.iram.fr/IRAMFR/GILDAS/}. The relative intensities of the hyperfine components given by Womack et al.~(\cite{wom}) were assumed. With the exception of Per~6, the line optical depths could be determined with at least 3.5 $\sigma$ in all objects (see Tab.~\ref{n2h}).  T$\rm _{ex}$ was estimated using the equation
$$
\rm T_{MB}=(J(T_{ex})-J(T_{BG}))\cdot (1-e^{-\tau})
$$
where T$\rm_{ex}$ is the excitation temperature, T$\rm_{BG}$ the cosmic background temperature (2.7~K) and 
$\rm J(T) = [h\nu /k] / [exp(h\nu/kT)-1]$.
In most cases, the excitation temperature of N$_2$H$^+$ is lower than the kinetic gas temperature listed in Jijina et al.~(\cite{jij}) and Hatchell~(\cite{nh3}), most likely due to line of sight average volume densities of the emitting gas lower than the critical density of the observed transitions ($\simeq$ 10$^{5}$ cm$^{-3}$ for N$_2$H$^+$(1--0)).
The factor by which T$\rm _{ex}$ is lower than T$\rm _{kin}$ varies from source to source from close to 1 (L~1448~IRS~3) to 0.43 (HH~211). However, these excitation temperatures are typically 2-3~K higher than those in the prestellar cores observed by Crapsi et al.~(\cite{crap}). Another contrast with the prestellar cores, is that none of the lines in our sample show highly non-Gaussian profiles.

The N$_2$H$^+$~1-0 line widths found in our protostellar sample are significantly larger than the ones in prestellar cores. The average line width for our sample is 0.61~kms$\rm ^{-1}$, whereas Crapsi et al.~(\cite{crap}) found an average line width of 0.26~kms$\rm ^{-1}$.  The line widths reported by Roberts \& Millar~(\cite{RM}) are slightly broader than ours (0.85~kms$^{-1}$), but this difference might be due to the lower spatial resolution of their observations. This suggests that nonthermal motions are not only concentrated nearby the protostar, but that they pervade the whole associated envelope,increasing outward as commonly found in molecular cloud cores (e.g. Fuller \& Myers \cite{fm92}).

N$_2$D$^+$ was observed in three different transitions (J=1-0, J=2-1 \& J=3-2). All three lines were detected in six objects. The ground-state transition, N$_2$D$^+$~1-0, was detected in only seven sources, but the S/N ratio of four of these is too low to obtain a reliable $\tau$ from hyperfine component fitting. Only in L~1448~IRS~3 and Barnard~1-b is the line strong enough; the HFS method yielded $\rm \tau = 1.3\pm 0.4~and~0.9\pm 0.2$, respectively. The reason for the rarity of N$_2$D$^+$~1-0 detections is that the Einstein-A coefficient, i.e. the transition probability, increases with $\rm \sim J^3$, and thus the A-coefficient for the J=1-0 line is ten times lower than the A-coefficient of the J=2-1 line. Furthermore, the relatively low energies ($E=J(J+1)\cdot 1.85~K$) and the higher degeneracies  of higher rotational states lead to an efficient population of those states under conditions present in protostellar cores (n$\rm \simeq 10^5~cm^{-3}$, T=10-20~K). 

The 2-1 line is the most frequently detected N$_2$D$^+$ line. Only two objects, Per~6 and L~1527, do not show emission in J=2-1. These two objects also show no emission in the other N$_2$D$^+$ lines. In most cases, the J=2-1 lines are optically thin, so that $\tau$ can be determined by fitting the hyperfine components in only five cases.

The N$_2$D$^+$~3-2 line is detected in 15 out of the 20 cores. Because the satellite lines get weaker with increasing J, and due to the higher noise-level of most of these data, it was not possible to derive reliable optical depths. 
 
The integrated intensities of the N$_2$H$^+$~1-0 lines from the cores in our sample are comparable to the intensities found in prestellar cores (Crapsi et al.~\cite{crap}), but the optical depths in the protostellar cores are typically lower by a factor of three. The integrated intensity and optical depth of N$_2$D$^+$~2-1 are approximately the same in pre- and protostellar cores, but N$_2$D$^+$~3-2 is significantly stronger in the latter ($\sim 0.3$~Kkms$^{-1}$ and $\sim 1.42$~Kkms$^{-1}$, respectively).

\subsection{C$^{18}$O \& HCO$^+$}

The C$^{18}$O~1-0 and HCO$^+$~3-2 lines were detected in all 20 sources of our sample. The line parameters are listed in Tab.~\ref{coh}. The line profile of C$^{18}$O, which is assumed to be optically thin, looks roughly Gaussian. The only two exceptions are L~1455~A1 and IRAS~03282, which show more complex line shapes, indicating that these sources may contain several velocity components. The C$^{18}$O line width (FWHM) is quite narrow, typically $<$1 km/s. Such narrow line widths indicate that the bulk of the molecular gas is cold, and turbulence is low. The column densities we calculated for C$^{18}$O (see below), indicate that the C$^{18}$~1-0 line is optically thin in these objects. 
            
The line shape of HCO$^+$~3-2 deviates clearly from a Gaussian profile. In 12 objects double-peaked profiles appear, with minima at the position of the optically thin C$^{18}$O emission, a clear indication for self-absorption. Asymmetries in the line shape indicate a systematic velocity pattern, which is probably caused by infall motions.

Gregersen et al.~(\cite{greg}) observed several young protostars in HCO$^+$~3-2 using the Caltech Submillimeter Observatory (CSO). They found line shapes that show similar asymmetries to those we detect (see section~\ref{anav}), but the intensities of the lines they observed are significantly lower. This discrepancy is most likely due to the larger beam size (32.5$''$) of the CSO at 268~GHz. For three sources common to the Gregersen et al.~(\cite{greg}) study and ours, we determined beam filling factors based on the HCO+ 3-2 lines. We used a CSO beam efficiency of 69.8\% at 269~GHz to convert their T$\rm_A^*$ to T$\rm_{MB}$. Assuming that HCO$^+$~3-2 is emitted from a small region at the centre of the core, as suggested by our models (see section~\ref{strat}), we estimated the size of this region by comparing the T$_{MB}$ ratio to the ratio of the beam size of the two observations. The derived filling factors for the CSO observations were between 0.26 and 0.36, yielding a diameter of the warm nucleus of approximately 4000~AU. Contributions of the error beams to the IRAM observations have been neglected here, but due to the small size of the HCO$^+$ emitting region, their contribution is $<1\%$.

\begin{table}[ht]
\caption{Observational parameters of the C$^{18}$O and HCO$^+$ lines. Integrated HCO$^+$ intensities from double-peaked spectra are indicated with an $^{*}$.} \label{coh}
\begin{center}
\begin{tabular}{lccccc}
\hline
  Source & W [Kkm/s]  & $\rm \Delta v$ [km/s] & W [Kkm/s]  \\
  & (C$^{18}$O) & (C$^{18}$O) & (HCO$^+$) \\
\hline
\hline
L 1448 IRS 2      & 3.23$\pm 0.05 $ & 0.77$\pm 0.01 $  & 13.0$\pm 0.3\rm^{*}$\\
L 1448 IRS 3      & 5.75$\pm 0.07 $ & 1.22$\pm 0.01 $  & 29.0$\pm 0.3$\\
L 1448 C          & 3.36$\pm 0.06 $ & 1.11$\pm 0.01 $  & 20.6$\pm 0.3\rm^{*}$  \\
L 1455 A1         & 1.65$\pm 0.06 $ & 1.03$\pm 0.01 $  &  8.8$\pm 0.3$   \\
L 1455 A2         & 0.90$\pm 0.05 $ & 0.80$\pm 0.01 $  &  6.5$\pm 0.3$   \\
Per 4 S           & 2.84$\pm 0.06 $ & 0.81$\pm 0.01 $  &  8.7$\pm 0.3$     \\
Per 4 N           & 1.30$\pm 0.04 $ & 0.44$\pm 0.01 $  &  5.3$\pm 0.3$    \\
Per 5             & 1.79$\pm 0.05 $ & 0.69$\pm 0.01 $  & 15.0$\pm 0.3$     \\
Per 6             & 1.52$\pm 0.05 $ & 0.74$\pm 0.01 $  &  6.1$\pm 0.3\rm^{*}$   \\
IRAS 03282        & 1.95$\pm 0.07 $ & 1.47$\pm 0.01 $  & 12.8$\pm 0.3\rm^{*}$ \\
Barnard 1-B       & 8.09$\pm 0.08 $ & 1.59$\pm 0.01 $  & 14.0$\pm 0.3\rm^{*}$ \\
HH 211            & 5.19$\pm 0.06 $ & 0.81$\pm 0.01 $  & 17.4$\pm 0.3\rm^{*}$    \\
Barnard 5 IRS 1   & 3.99$\pm 0.06 $ & 1.00$\pm 0.01 $  &  9.3$\pm 0.3\rm^{*}$ \\
L 1527            & 4.12$\pm 0.06 $ & 0.65$\pm 0.01 $  & 13.0$\pm 0.3\rm^{*}$    \\
L 483             & 5.12$\pm 0.06 $ & 0.81$\pm 0.01 $  & 22.8$\pm 0.3\rm^{*}$  \\
SMM 5             & 8.66$\pm 0.08 $ & 1.57$\pm 0.01 $  & 10.7$\pm 0.3$          \\
L 723             & 2.45$\pm 0.08 $ & 1.29$\pm 0.01 $  & 12.2$\pm 0.3\rm^{*}$         \\
L 673 A           & 2.31$\pm 0.07 $ & 0.92$\pm 0.01 $  & 12.8$\pm 0.3$        \\
B 335             & 2.34$\pm 0.04 $ & 0.44$\pm 0.01 $  &  3.6$\pm 0.3\rm^{*}$           \\
L 1157            & 1.57$\pm 0.05 $ & 0.61$\pm 0.01 $  & 13.4$\pm 0.3\rm^{*}$         \\
\hline
\end{tabular}
\end{center}
\end{table}    

\section{Analyses}

In this section, we calculate the N$_2$D$^+$/N$_2$H$^+$ ratios and correlate them with physical parameters of the protostars that might influence deuterium fractionation. Some of these parameters, such as dust temperature and bolometric luminosity, can be used to quantify the evolutionary stage of the protostar. Each parameter will be discussed in individual subsections.        

\subsection{N$_2$H$^+$ \& N$_2$D$^+$ volume densities, column densities and deuterium fractionation}\label{ana1}

To derive the deuterium enhancement, one calculates the ratio of the column densities of a hydrogen-bearing species and its deuterated isotopologue.  Many common C-bearing  molecules (e.g. HCO$^+$ and H$_2$CO) suffer from depletion in cold and dense environments (see Tafalla et al.~\cite{taf06}), and thus the deuterated species trace slightly warmer regions of the core, where the parent molecules are not depleted. To properly determine the deuterium fractionation of an entire core, one has to observe molecules that are less affected by depletion, such as N$_2$H$^+$ and NH$_3$. Another benefit of observing N-bearing molecules is that their emission lines split into hyperfine components due to the non-zero nuclear spin of nitrogen, enabling optical depths to be measured.  

To determine molecular column densities, we used the N$_2$H$^+$ 1-0 and N$_2$D$^+$ 2-1 lines. We chose these lines for three reasons. First, N$_2$D$^+$~2-1 is detected in most of the sources. Second, the 2-1 transition of N$_2$D$^+$ is observed with the highest S/N ratio. Third, the ratio of the N$_2$D$^+$~2-1/N$_2$H$^+$~1-0 lines permits a direct comparison with previous studies (Crapsi et al.~\cite{crap}, Fontani et al.~\cite{fon}).
% so we decided to use the same method as they did.
The column densities were derived (Tab.~\ref{D/H}) assuming a constant excitation temperature, as determined in section~\ref{res} (CTEX-method, for details see the appendix of Caselli et al.~\cite{cas2}). The excitation temperature of N$_2$H$^+$ can be determined in all objects but Per 6 (Tab.~\ref{n2h}), where the N$_2$H$^+$ line is optically thin. For Per~6, we assumed an excitation temperature of 8~K, which is the mean excitation temperature of the sample.  
For N$_2$D$^+$, the situation is different, since the  N$_2$D$^+$ 2-1 line is optically thin in most objects.  In the few sources where $\tau$ could be determined, the excitation temperatures of N$_2$H$^+$ and N$_2$D$^+$ were approximately the same. In all cases where the excitation temperature of N$_2$D$^+$ was unknown, we assumed that it was the same as for N$_2$H$^+$. The estimated errors on the column densities are of the order of 15\% and 20\% for N$_2$H$^+$ and N$_2$D$^+$, respectively. As expected, the N$_2$D$^+$/N$_2$H$^+$ ratios vary significantly from object to object (Tab.~\ref{D/H}). The extreme values are $<0.029$ for Per~6 and 0.27 for HH~211. For most objects, the ratio is below 0.1.

Determining T$_{ex}$ and the column densities as described above, we implicitly assume a unity beam filling factor for the N$_2$H$^+$ and N$_2$D$^+$ observations. Because dark cloud cores are typically larger than the beam size of our observations (e.g. Caselli et al.~\cite{cormap}) and the fact that the kinetic temperature of these cores is only $\sim 1.5$ times  the excitation temperature of N$_2$H$^+$ indicates that the filling factor is indeed close to one, and certainly $>0.67$. Using the lower limit for the filling factor would lower the N$_2$H$^+$ column density by 40\% and the N$_2$D$^+$ column density by 55\%.  Therefore, the derived N$_2$D$^+$/N$_2$H$^+$ ratios could be up to 25\% lower. The general trend, however, which we will discuss in the next section, remains.

The N$_2$D$^+$ column density was also calculated using the (3--2) line, where detected, following the same procedure described above. In 13 out of 15 objects, the values for the column density agree within the range of error. In L~1448~IRS3 and L~1157, the column densities determined using N$_2$D$^+$~3-2 are a factor of two higher than those determined using the 2-1 line. However, because of the lower S/N ratio of the N$_2$D$^+$~3-2 line, the column densities determined using the J=2-1 transition are more reliable. 

\begin{table}[ht]
\caption{N$_2$H$^+$ and N$_2$D$^+$ column densities.} \label{D/H}
\begin{center}
\begin{tabular}{lccc}
\hline
  Source & N(N$_2$H$^+$) & N(N$_2$D$^+$) & N(N$_2$D$^+$)/N(N$_2$H$^+$) \\
  & [cm$^{-2}$] & [cm$^{-2}$] \\
\hline
\hline
L 1448 IRS 2 & 3.1$\cdot 10^{13}$ & 1.3$\cdot 10^{12}$ & 0.042$\pm$0.014\\
L 1448 IRS 3 & 4.8$\cdot 10^{13}$ & 3.9$\cdot 10^{12}$ & 0.081$\pm$0.010\\
L 1448 C & 2.4$\cdot 10^{13}$ & 2.5$\cdot 10^{12}$ &0.104$\pm$0.017\\
L 1455 A1 & 9.2$\cdot 10^{12}$ & 6.3$\cdot 10^{11}$ & 0.068$\pm$0.043\\
L 1455 A2 & 1.4$\cdot 10^{13}$ & 9.6$\cdot 10^{11}$ & 0.069$\pm$0.029\\
Per 4 S & 7.9$\cdot 10^{12}$ & 1.4$\cdot 10^{12}$ & 0.177$\pm$0.032\\
Per 4 N & 7.1$\cdot 10^{12}$ & 7.8$\cdot 10^{11}$ & 0.110$\pm$0.057\\
Per 5 &  1.2$\cdot 10^{13}$ & 1.0$\cdot 10^{12}$ & 0.083$\pm$0.033\\
Per 6 &  6.1$\cdot 10^{12}$ & $<$1.8$\cdot 10^{11}$ & $<$ 0.029\\
IRAS 03282 & 1.6$\cdot 10^{13}$ & 3.5$\cdot 10^{12}$ & 0.219$\pm$0.026 \\
Barnard 1-B & 4.5$\cdot 10^{13}$ & 8.1$\cdot 10^{12}$ & 0.180$\pm$0.009\\
HH 211 &   1.7$\cdot 10^{13}$ & 4.6$\cdot 10^{12}$ & 0.271$\pm$0.024\\
Barnard 5 IRS 1 &  1.6$\cdot 10^{13}$ & 5.4$\cdot 10^{11}$ & 0.034$\pm$0.025\\
L 1527 &  7.2$\cdot 10^{12}$ & $<$ 2.5$\cdot 10^{11}$ & $<$0.034\\
L 483 & 3.3$\cdot 10^{13}$ & 1.9$\cdot 10^{12}$ & 0.057$\pm$0.012\\
SMM 5 & 2.9$\cdot 10^{13}$ & 2.3$\cdot 10^{12}$ & 0.079$\pm$0.019  \\
L 723 &  8.8$\cdot 10^{12}$ &  4.0$\cdot 10^{11}$ & 0.045$\pm$0.040  \\
L 673 A &   1.3$\cdot 10^{13}$ & 4.3$\cdot 10^{11}$ & 0.033$\pm$0.031   \\
B 335 &   1.4$\cdot 10^{13}$ & 6.0$\cdot 10^{11}$ & 0.043$\pm$0.029     \\
L 1157 &  1.1$\cdot 10^{13}$ & 5.0$\cdot 10^{11}$ & 0.045$\pm$0.036   \\
\hline
\end{tabular}
\end{center}
Estimated errors on the column densities are about 15\% and 20\% for N$_2$H$^+$ and N$_2$D$^+$, respectively.
\end{table}

The N$_2$H$^+$ and N$_2$D$^+$ column densities derived here are of the same order as those found by Crapsi et al.~(\cite{crap}) for prestellar cores ($\rm 10^{13}~cm^{-2}$ and $\rm 10^{12}~cm^{-2}$ for N$_2$H$^+$ and N$_2$D$^+$, respectively).This result is somewhat unexpected, because the optical depth of N$_2$H$^+$ is about three times lower in protostars than in prestellar cores. However, the higher excitation temperatures and broader line widths of N$_2$H$^+$ in protostellar envelopes (over-)compensate for the lower optical depth.
The range of values for the N$_2$D$^+$/N$_2$H$^+$ ratio is also similar. The average ratio in our sample of Class~0 protostars is 0.097, and 20\% of the objects have a ratio larger than 0.15. For starless cores, the average ratio is 0.106, and 18\% of the sources have ratios larger than 0.15. However, the N$_2$D$^+$/N$_2$H$^+$ ratio in both protostellar and prestellar cores is significantly larger than the ratio in high mass star-forming regions (always smaller than 0.02, Fontani et al.~\cite{fon}), although values as large as 0.1 have been measured with interferometric observations (Fontani et al.~\cite{fcb08}).

For three objects in our sample, L~1448~C, HH~211, and L~1527, the N$_2$D$^+$/N$_2$H$^+$ ratio was also determined by Roberts \& Millar~(\cite{RM}). In general, they find somewhat higher N$_2$D$^+$/N$_2$H$^+$ ratios than we do. In L~1448~C, their ratio is 1.63$\pm 0.4$ times larger. In HH~211, their ratio is 1.14$\pm 0.18$ times larger. For L~1527, we report an upper limit of 0.034, whereas Roberts \& Millar found a ratio of 0.06. The fact that Roberts \& Millar~(\cite{RM}) found similar N$_2$D$^+$/N$_2$H$^+$ ratios at much lower spatial resolution indicates that the deuterium fractionation occurs, as expected, in the cold, extended envelope of the protostellar core. If deuterium fractionation occurred only in the central region, one would expect beam dilution to yield lower  N$_2$D$^+$/N$_2$H$^+$ ratios at lower resolutions.

We also estimated the volume densities of the cores. Because the N$_2$D$^+$ lines show only a single velocity component, we used the ratios of the integrated intensities of these lines instead of the T$_{MB}$ ratios. Thus we obtained a higher signal to noise ratio. Further we assumed an isothermal cloud with a kinetic temperature determined from NH$_3$ observations (Tab.~\ref{n2h}). The volume density was estimated by fitting the N$_2$D$^+$ J=3-2/J=2-1 ratio, using a radiative transfer code (Pagani~\cite{MCn}, see section~\ref{codes}). We derived values between $\rm 5\cdot 10^6~cm^{-3}$ and $\rm 10^7~cm^{-3}$. Only IRAS~03282 shows a lower density ($\rm 6\cdot 10^5~cm^{-3}$). 
However, temperature and density decline with distance from the protostar, so lower spatial resolution observations trace lower average temperatures, densities and (assuming spherical symmetry) lower column densities. Therefore, the N$_2$D$^+$~3-2 lines trace, on average, warmer and denser gas than the N$_2$D$^+$~2-1 lines, and the densities obtained from the ratios are upper limits. For six sources, we also determined densities using the N$_2$D$^+$ J=2-1/J=1-0 ratios. We were not able to find a density which would reproduce all three lines simultaneously. The density values derived from the J=2-1/J=1-0 line ratio  are about a factor 10 lower than those determined from the J=3-2/J=2-1 ratio. This difference is again most likely arises because of the lower spatial resolution of the lower frequency observations.

\subsection{Dust Temperature}\label{Tdustsec}

In general, T$\rm _{Dust}$ is expected to rise as a protostar evolves, but there is no exact model of dust temperature as a function of time. As a result, connecting dust temperature to specific protostellar ages depends on which model is used (see Froebrich~\cite{fro} and references therein). 

All of our sources, with the exceptions of Per~4~S and SMM~5, have infrared counterparts identified in the IRAS point source or faint source catalogues. In addition, millimetre and submillimetre continuum data are available for many of the sources (see Tab.~\ref{Tempd}). 
We fitted these data to derive values for dust temperature (T$\rm _{Dust}$) and emissivity ($\beta$) using equation (1) from Kramer et al.~(\cite{car}): 
$$
\rm F_\nu\propto \nu ^{3+\beta} \cdot \frac{1}{e^{h\nu /kT_{Dust}}-1}
$$
where F$_\nu$ is the continuum flux at frequency $\nu$.
For 15 out of 16 sources for which more than two data points are available, a value of $\beta$=1 yielded the best fit. Only for Barnard~5~IRS1 does an index of $\beta$=0 (i.e. black-body radiation) yield a better result. In the three cases where data at only two frequencies were available, we assumed $\beta$=1. This value is lower than the typical emissivity index of $\beta =2$ for molecular clouds (e.g. Ward-Thompson et al.~\cite{war}, Visser et al.~\cite{vis}). We propose that the lower emissivity indices found in the Class~0 sources are a result of strong temperature gradients and not inherent differences in the dust grains themselves.
In fact, the cold envelopes are still bright at millimetre wavelengths, whereas practically all radiation emitted at $\rm\lambda <  100\mu m$ comes from the hot core. Therefore, the mass of dust radiating at longer wavelengths is much greater than the mass of dust radiating at shorter wavelengths, and the spectrum from the entire core gets flattened. 
We modelled the continuum emission using an emissivity index of two (see section~\ref{model}), but got as index for the whole modelled cloud a $\beta$-value of one.

\begin{table*}[ht]
\caption{Continuum observations and derived properties (dust temperature, C$^{18}$O column density, CO depletion factor) for sources in our survey. The errors in the dust temperatures are $\sim 2$~K, and the C$^{18}$O column densities are accurate to 10~\%. } \label{Tempd}
\begin{center}
\begin{tabular}{lccccccc}
\hline
  Source & IRAS & F(60$\mu$m) & F(100$\mu$m) & Wavelength of additional data  & T$\rm _{Dust}$ & N(C$^{18}$O) & f$\rm _D$(CO) \\
  & & [Jy] & [Jy] & [$\mu$m]  & [K] & [cm$^{-2}$] & \\
\hline
\hline
L 1448 IRS 2 & 03222+3034 & 14.3 & $<$169 & 450,~850,~1100$^{1,3}$   & 27           & 6.07$\cdot 10^{15}$ & $2.6\pm 0.6$   \\
L 1448 IRS 3 & 03225+3034 & 25.1 & 169& 1100$^{3}$  & 27                         & 1.08$\cdot 10^{16}$ & $4.4\pm 1.0$ \\ 
L 1448 C &   F03226+3033 & 53.4 & 355 & 450,~850,~1100$^{2,3}$   & 32                 &  7.18.19$\cdot 10^{15}$ & $2.8\pm 0.6$      \\
L 1455 A1 &  03245+3002 & 47.1 & 93.6 & 1100,~450,~850$^{3,5}$   & 37             &  3.53$\cdot 10^{15}$ & $1.5\pm 0.4$ \\
L 1455 A2 &  F03247+3001 & 66.4 & 206 & - &30                              &  1.83$\cdot 10^{15}$ & - \\
Per 4 S &   -  & - & - & 1100$^{3}$  & -                                         & 3.18$\cdot 10^{15}$ & - \\
Per 4 N &   03262+3123 & 1.07 & $<$12.4 & 1100$^{3}$ &23                        &  2.18$\cdot 10^{15}$ & $2.4\pm 0.11$\\
Per 5 &    03267+3128 & 1.88 & $<$10.4 & 1100$^{3}$  & 26                         &  3.28$\cdot 10^{15}$ & $1.6\pm 0.4$\\
Per 6 &    03271+3013 & 7.53 & 8.19 & - &45                                 &   4.28$\cdot 10^{15}$ & -\\
IRAS 03282 & 03282+3035 &2.33 & 13.9 & 450,~850,~1100$^{2,3}$  & 23              &  3.28$\cdot 10^{15}$ & $3.9\pm 0.9$\\
Barnard 1-B & 03301+3057 & $<$6.18 & 35.5 & 1100,~850$^{3,4}$  & 18             & 1.16$\cdot 10^{16}$ & $2.9\pm 0.7$\\
HH 211 &  03407+3152 & $<$2.06 & $<$30.8 & 1100,~450,~850$^{3,5}$  & $<$21        &   8.21$\cdot 10^{15}$ & $2.9\pm 0.7$\\
Barnard 5 IRS 1 & 03445+3242 & 15.5 & 15.4 & 1100$^{3}$  & 50                     & 1.23$\cdot 10^{16}$ & $0.3\pm 0.09$\\
L 1527 &  04368+2557 & 17.8 & 73.3 & 450,~850,~1300$^{2}$  & 27                   &  7.75$\cdot 10^{15}$ & $2.3\pm 0.5$\\
L 483 &  18148-0440  &89.1 & 166 & 450,~850$^{2}$  & 34                           &  1.15$\cdot 10^{16}$ & $2.1\pm 0.5$\\
SMM 5 &  -     &-&-    &-& -                                                & 9.70$\cdot 10^{15}$ & - \\
L 723 &  19156+1906 & 6.93 & 20.7 & 450,~850$^{2}$   & 25                   &  4.36$\cdot 10^{15}$ & $2.9\pm 0.7$\\
L 673 A & 19180+1114 &2.55 & 10.6 &  -&28                                      & 4.47$\cdot 10^{15}$ & -\\
B 335 & 19345+0727  & 8.3 & 42  & 450,~850,~1300$^{2}$  &27                        &   4.41$\cdot 10^{15}$ & $3.3\pm 0.8$ \\
L 1157 & 20386+6751  &10.9 & 43.5  & 1300,~450,~850$^{2,5}$  &30                    & 3.19$\cdot 10^{15}$ & $4.2\pm 1.0$\\
\hline
\end{tabular}
\end{center}
$^1$O'Linger et al.~(\cite{oli}), $^2$Shirley et al.~(\cite{shi}), $^3$Enoch et al.~(\cite{eno}), $^4$Hatchell et al.~(\cite{hcon}), $^5$Froebrich et al.~(\cite{fro2}).
\end{table*}

\begin{figure}[ht]
   {\resizebox{8cm}{!}
            {\includegraphics{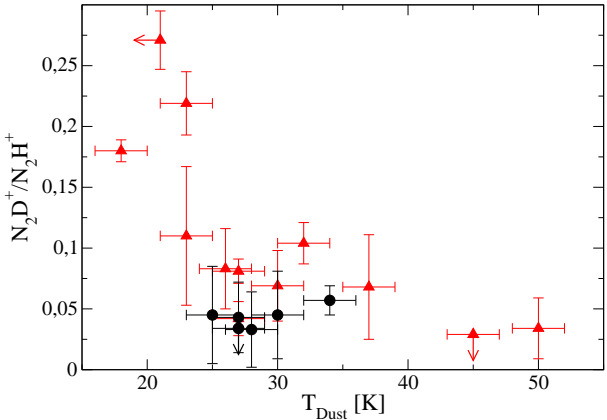}}}    
     \caption{Deuterium fractionation (N$_2$D$^+$/N$_2$H$^+$ column density ratio) versus dust temperature. The largest N$_2$D$^+$/N$_2$H$^+$ ratios are seen at the lowest temperatures. As T$\rm _{Dust}$ increases, the deuterium fractionation declines. The objects marked with triangles are located in the Perseus cloud.} \label{Td}
      \end{figure}  
      
The dust temperatures derived from fitting the continuum data range from approximately 20~K up to 50~K (see Tab.~\ref{Tempd}). These temperatures are clearly higher than the excitation temperatures of N$_2$H$^+$, as well as the kinetic gas temperatures (Tab.~\ref{n2h}). 
The difference is likely due to the fact that the dust emission from warmer regions is stronger than from cold dust, and thus the obtained average temperature is higher than the gas temperature traced by N$_2$H$^+$.

In Fig.~\ref{Td}, the  N$_2$D$^+$/N$_2$H$^+$ column density ratio is plotted versus the dust temperature. All objects with N$_2$D$^+$/N$_2$H$^+$ ratio $>$0.15 have T$\rm _{Dust}<$ 25~K. At higher dust temperatures, the deuterium fractionation declines, reaching an average value of approximately 0.05 at 30~K.

\subsection{CO freeze-out}\label{Cod}

Because CO is one of the main destroyers of H$_3^+$ and deuterated isotopologues, the progenitors of N$_2$D$^+$, the N$_2$D$^+$/N$_2$H$^+$ ratio should be sensitive to CO depletion (e.g. Dalgarno \& Lepp~\cite{dl84}, Roberts \& Millar~\cite{rob}).
The CO depletion factor, f$\rm _D$, is defined as
$$ 
\rm f_D = x_{can}/x_{obs}
$$
where x$\rm_{can}$ is the canonical ISM abundance ratio of CO. In this work, C$^{18}$O has been observed, so that $\rm x_{can} = \rm [C^{18}O]/[H_2]=1.7\cdot 10^{-7}$ (Frerking et al.~\cite{frer}). The total number of hydrogen molecules within the beam is determined by the $\sim 1$~mm continuum emission (1.3~mm, 1.1~mm, or 850~$\mu$m), using the equation 
$$
\rm {\cal{N}}_{H_2}=\frac{F_{\nu}\cdot d^2}{\kappa_{\nu}\cdot B_\nu (T_{Dust})}\cdot\frac{1}{\mu}
$$  
(Terebey et al.~\cite{ter}), where F$_\nu$ is the continuum flux, d is the distance to the object, B$_\nu $(T$\rm _{Dust}$) is the Planck function, T$\rm _{Dust}$ is the dust temperature determined in the previous subsection, and $\mu$ is the mass of an H$_2$ molecule. For the opacity per unit mass, $\kappa_\nu$, we used a value of 0.005 $\rm cm^2/g$ at a wavelength of 1.2~mm (Andr\' e et al.~\cite{kappa}). To calculate $\kappa_\nu$ at other wavelengths, we express the relation between opacity and wavelength as a power law
$$
\rm  \kappa _\lambda = \kappa _0 \left( \frac{\lambda _0}{\lambda}\right) ^\beta.
$$
We used $\beta=1$ (zero for Barnard~5~IRS1), as suggested by the fits to the continuum data (see section~\ref{Tdustsec}). The values for $\kappa$ make the largest contributions to the uncertainties in the H$_2$ column densities. Some authors use values for $\kappa$ twice as high as ours (e.g. Froebrich~\cite{fro}).
%The uncertainty of T$\rm _{Dust}$ arises, because we use a single temperature obtained by a fit to observations at wavelength from 60~$\mu$m to 1~mm, although temperature gradients are expected in protostars. Whether the obtained temperature is representative for dust emission at $\lambda\sim 1$~mm is not clear.

To compare the $\sim 1$~mm continuum data with our C$^{18}$O~1-0 observations, we have to estimate the continuum flux density within a 22$''$ beam. The fact that the angular diameters of the protostellar envelopes are always larger than the beam is an important issue in this conversion. To compare data taken with different beam sizes, we follow the method of Terebey et al.~(\cite{ter}). They showed that the flux of a dark cloud with a density distribution $\rm\rho = r^{-p}$ and a temperature distribution $\rm T=r^{-q}$ can be described by:
$$
\rm F_\nu \propto d^{1-(p+q)}\cdot \theta ^{3-(p+q)}
$$
where d is the distance and $\theta$ is the beam size. In their sample of mostly Class~0 protostars, they determined 1.7 to be an appropriate value for (p+q). We used this value to scale the flux from the continuum observations to the C$^{18}$O beam size.
%The CO depletion factor is calculated with the above formulae. 
 
We determined the C$^{18}$O column density by using the CTEX-method and by assuming that C$^{18}$O emission is optically thin.  
Since the critical density of C$^{18}$O is 60 times lower than that for N$_2$H$^+$, we expect that $\rm T_{ex}(C^{18}O)$ will be higher than $\rm T_{ex}(N_2H^+)$ and comparable to the kinetic gas temperature or even slightly higher, due to freeze-out of CO in the cold parts of the cloud. Since the beam sizes of the NH$_3$ observations which were used to derive T$\rm _{kin}$ vary between  40$''$ and 88$''$, we adopted a constant excitation temperature of 15~K for all sources. This assumption introduces some error ($<$ 15\%) in the estimation of the depletion factor. The total number of C$^{18}$O molecules is then calculated from 
$$
\rm {\cal{N}}_{C^{18}O}= N_{C^{18}O}\cdot d^2\cdot\Omega
$$
where ${\cal{N}}\rm _{C^{18}O}$ is the total number of C$^{18}$O molecules within the beam, N$\rm _{C^{18}O}$ is the C$^{18}$O column density, d is the distance to the protostar, and $\Omega$ is the beam solid angle.

The derived depletion factors lie between 1 and 4 (Tab.~\ref{Tempd}). A value less than 1 is found toward Barnard 5 IRS 1 and this may be an indication that the C$^{18}$O canonical abundance is larger  in this source.  Indeed, the $^{13}$CO abundances measured toward Barnard 5 by Pineda et al. ~(\cite{pcg08}) are a factor of two larger than those measured by Frerking et al. ~(\cite{frer}) toward Taurus (CO abundance variations of about a factor of 2 are well known in star forming regions, e.g. Lacy et al.~\cite{lkg94}).  However, given that CO abundance measurements are not individually available for all the sources, the C$^{18}$O canonical abundance has been kept the same for the whole sample.

\begin{figure}[ht]
   {\resizebox{8cm}{!}
            {\includegraphics{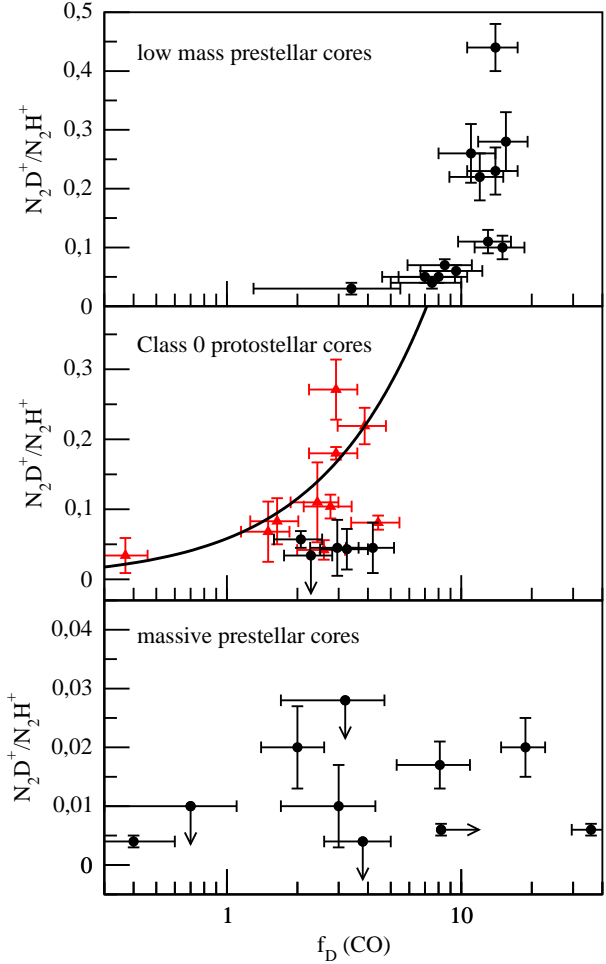}}}    
     \caption{N$_2$D$^+$/N$_2$H$^+$ column density ratio versus CO depletion factor for prestellar cores (upper panel, Crapsi et al.~\cite{crap}), Class~0 protostars (middle panel, this work) and massive prestellar cores (lower panel, Fontani et al.~\cite{fon}). The protostellar objects marked with triangles are located in the Perseus cloud. Among the Perseus objects, f$\rm _D$(CO) and the deuterium fraction appear correlated. The solid curve in the middle panel is the result of fitting the sub-sample of Perseus cores with a formula adopted from Caselli et al.~(\cite{cw98}, see text).} \label{fdplot}
      \end{figure}

In the central panel of Fig.~\ref{fdplot}, we show the correlation between CO freeze-out and deuterium fractionation in our sample of protostars. In the top panel, we show the same correlation found in low-mass prestellar cores (Crapsi et al.~\cite{crap}), and in the bottom panel, the correlation for massive prestellar cores (Fontani et al.~\cite{fon}). 
The correlation determined from our sample looks very similar to the correlation found for low-mass prestellar cores. For small depletion factors (f$\rm _D$(CO)$<3$), the deuterium fractionation is low as well ($< 0.1$). For depletion factors greater than three, the N$_2$D$^+$ /N$_2$H$^+$ ratio rises quickly until it reaches 0.25. The only difference between the protostellar and prestellar cores is that in the prestellar cores, the deuterium fractionation stays low until the depletion factor is $\sim 10$.  This behavior reflects the fact that CO is highly depleted in the center of prestellar cores, whereas it is not depleted in the centers of Class 0 sources. In high mass prestellar cores, there is no dependence of the deuterium fractionation on f$\rm _D$(CO). Since the high mass prestellar cores in the Fontani et al.~\cite{fon} sample are at larger distances (a few kpc), these observations might be affected by non-depleted, non-deuterated gas along the line of sight.

The sources that are {\em not} located in the Perseus cloud show lower deuterium fractionation than Perseus sources with the same CO depletion factor. One explanation for this difference is that the Perseus sources are embedded in a dense environment ($\rm n_{H_2}\sim 10^4~cm^{-3}$, Kwon et al.~\cite{dens1448}), whereas many of the other cores are isolated. Since CO is not expected to be frozen out in the embedding material, the CO depletion factors of the Perseus cores could be underestimated.  
However, systematic errors arising from the different spatial resolutions and different wavelengths of the continuum observations cannot be ruled out. The Perseus cores were observed at $1100~\mu m$ with a resolution of 31$''$ (Enoch~\cite{eno}); the other cores were observed at $1300~\mu m$ with a resolution of 40$''$ (Shirley et al.~\cite{shi}).
The Perseus source which does not appear to agree with the correlation is L~1448~IRS~3. It has a  CO depletion factor of 4.4 but a low deuterium fractionation (0.081). This source is known to consist of multiple cores with several strong outflows (e.g. Volgenau et al.~\cite{nik},  Looney et al.~\cite{loo}), but it is not clear how this activity can decrease the deuterium fractionation while maintaining a large amount of CO freeze-out.

The correlation among the sub-sample of Perseus sources is quite good, whereas the non-Perseus sources do not show this functional dependence. We made a least-squares fit of the Perseus data to the equation
$$
\rm N_2D^+/N_2H^+=\frac{c_1}{(c_2+k_{CO}\cdot x_{can}(CO)/f_D(CO))},
$$
which we adopted from expression (1) in Caselli et al.~(\cite{cw98}).
In this equation, $\rm c_1=\rm 1/3\cdot k_f\cdot x(HD)$, (where $\rm k_f$ is the formation rate of H$_2$D$^+$ and x(HD) is the HD abundance), c$_2$ is the electron abundance (x(e)) times the dissociative recombination rate of H$_2$D$^+$, $\rm k_{co}$ is the destruction rate of H$_2$D$^+$ by CO, and $\rm x_{can}(CO)$ is the canonical CO abundance. The equation neglects the effect of multi-deuterated species and also assumes that dissociative recombination is a significantly faster process than recombination onto dust grains. A more comprehensive chemical model will be used in Sect.~\ref{model}. The result of the least-squares fit is indicated by the solid curve in Fig.~\ref{fdplot}. The correlation coefficient is 0.71, indicating that the data agree fairly well with the expected theoretical fit. 
Using the reaction rates given by Caselli et al.~(\cite{cw98}) and assuming a temperature of 10~K, we determined the HD abundance and the electron abundance. The derived an HD abundance of $3\cdot 10^{-5}$ is, as expected, quite close to the cosmic D/H ratio. The electron abundance we find is $\sim 7\cdot 10^{-9}$, which is at the low end of the values found in dense cloud cores.   
However, due to the uncertainties in the temperatures, densities, and the determination of f$\rm _D$(CO) (see below), errors in the HD and electron abundances determined in this way may be as large as an order of magnitude. 

%A reason why the correlation of the deuterium fractionation with the CO depletion factor may be apparent in the Perseus sources may be because the Perseus sources all lie at approximately at the same distance. Furthermore, the continuum observations used for the determination of f$\rm _D$ were all made at 1100~$\mu$m, with the same beam width ($\theta = 31''$). As a result, systematic errors introduced by the correction for the beam width are identical for these sources and lead to a constant scaling error of the CO depletion factor. The non--Perseus sources were observed at different wavelengths and resolutions, which causes scatter in the N$_2$D$^+$/N$_2$H$^+$ versus f$\rm _D$ diagram. 

The errors in the depletion factor are of the order of 20~\%. These errors do not include the uncertainties in the excitation temperatures, dust temperatures and $\kappa$'s, which are the main sources of systematic errors in this analysis. A different $\kappa$ value, however, would just rescale f$\rm _D$, but it would not change the correlation itself. Such a rescaling would also be the effect of a different dust temperature, because f$\rm _D$(CO) was determined from continuum measurements around 1~mm, which is a wavelength clearly in the Rayleigh-Jeans regime.    
Rejecting the assumption of a constant average excitation temperature in all objects would change the correlation significantly.

If $\rm T_{ex}(C^{18}O)$ is lower than 15~K, we $under$estimate the depletion factor, if $\rm T_{ex}(C^{18}O)$ is greater than 15~K we $over$estimate it. But, if we assume that T$\rm _{ex}$(C$^{18}$O) is related to the dust temperature, i.e. objects with higher T$\rm _{Dust}$ have also higher C$^{18}$O excitation temperatures, and taking the correlation between the N$_2$D$^+$/N$_2$H$^+$ ratio found in the previous section into account, the high deuterium fractionation cores would may have greater f$\rm _D$(CO) values, and the low deuterium fractionation cores would have lower f$_D$(CO) values. Therefore the correlation would spread out further along the depletion factor axis.
   
\subsection{Bolometric Luminosity}\label{ana4}

The SED of a young stellar object changes as the central protostar heats up the surrounding gas and dust. Several measures have been developed to determine how far a protostar has evolved, including the bolometric temperature (Myers \& Ladd~\cite{tbol}), the ratio of the submillimetre to bolometric luminosity (Andr\' e et al.~\cite{Class0}), and the correlation between luminosity (L) and visual extinction (A$\rm _v$) (Adams~\cite{lav}). Another measure of the evolution of a protostar, developed by Saraceno et al.~(\cite{sar}), is the ratio of $\rm L_{BOL}~to~F_{1.3}^{D_0}$, the distance-normalised flux at 1.3~mm. As the protostar warms up, L$\rm _{BOL}$ should rise quickly, whereas $\rm F_{1.3}^{D_0}$, which is a measure for the envelope mass and less dependent on the temperature, should stay constant or decrease slightly. Hence, the ratio $\rm L_{BOL}/F_{1.3}^{D_0}$ increases as the protostar evolves.
% We choose this method, because  the required data were available in the literature for a significant fraction of our source sample. 

%The bolometric luminosities as well as the millimetre continuum data of several sources of our sample can be found in the literature.
To determine $\rm F_{1.3}^{D_0}$ values, we use fluxes at $\rm\lambda = 1.3~mm~(5~sources),~1.1~mm~(4~sources)~and~850~\mu m~(2~sources)$. The continuum fluxes at 1.1~mm and 850~$\mu$m were corrected using, as before, $\beta=1$ (or, in the case of Barnard~5~IRS~1, $\beta=0$). For the distance normalisation and the conversion of the different beam sizes, we used the equation from Terebey et al.~(\cite{ter}) described in the previous section, i.e.
$$
\rm F_{1.3}^{220~pc}=F_{1.3}^{obs.}\cdot(d/220~pc)^{0.7}\cdot(\theta/31'')^{-1.3},
$$
where $\rm F_{1.3}^{obs.}$ is the continuum flux, d is the distance to the object, and $\theta$ is the beam size. We normalised all observations to a distance of 220~pc and a beam size (FWHM) of 31$''$. From the errors given for the continuum fluxes and bolometric temperatures, we estimate an error of 30 \% in the calculated $\rm L_{BOL}/ F_{1.3}^{D_0}$ ratios. The resulting normalised fluxes and the bolometric luminosities for the cores are listed in Tab.~\ref{lbol}.

\begin{table}[ht]
\caption{Bolometric luminosity and distance-normalised flux of Class~0 protostars in our sample.} \label{lbol}
\begin{center}
\begin{tabular}{lccc}
\hline
  Source & L$\rm _{BOL}$~[L$_\odot$] & F$\rm _{1.3}^{220~pc}$  & L$\rm _{BOL}$/F$\rm _{1.3}^{220~pc}$\\
\hline
\hline
L 1448 IRS 2  & 6.0$^1$ & 0.80 & 7.5 \\
L 1448 C  & 8.3$^1$  & 0.53 & 15.6\\
L 1455 A1  & 9.7$^1$ & 0.19 & 50.2\\
IRAS 03282  & 1.2$^2$ & 0.32 & 3.8\\
Barnard 1-B  & 2.8 & 0.74 & 3.8\\
HH 211 &   3.6$^1$ &  0.61 & 5.9 \\
L 1527 & 1.9$^1$ &  0.38  & 5.0  \\
L 483  &  13.0$^2$ & 0.70 & 18.5   \\
SMM~5 & 3.6$^1$  &- & -\\
L 723  & 3.3$^2$ & 0.45 & 7.4   \\
B 335 &  3.1$^2$ & 0.45 & 6.9    \\
L 1157 &  5.8$^2$ & 0.68 & 8.6  \\
\hline        
\end{tabular}
\end{center}
$^1$Froebrich~(\cite{fro}), $^2$Shirley et al.~(\cite{shi}).
\end{table}

\begin{figure}[ht]
   {\resizebox{8cm}{!}
            {\includegraphics{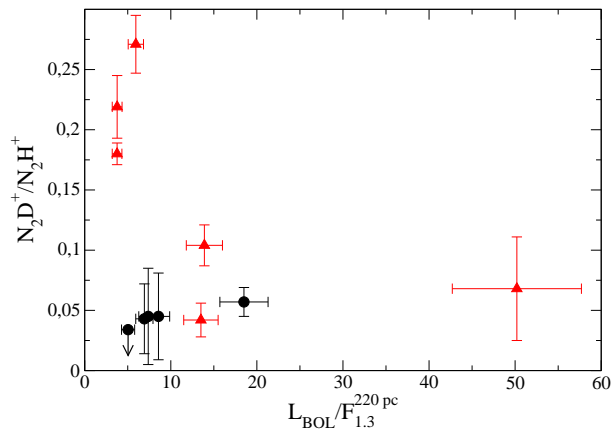}}}    
     \caption{N$_2$D$^+$/N$_2$H$^+$ versus $\rm L_{BOL}/ F_{1.3}^{D_0}$. All objects  with high deuterium fractionation have low L$\rm_{BOL}/F_{1.3}^{D_0}$ ratios and are likely to be young objects. L~1527, which has one of the lowest L$\rm_{BOL}/F_{1.3}^{D_0}$ ratios, also has a quite upper limit on the N$_2$D$^+$/N$_2$H$^+$ ratio.   } \label{ll}
      \end{figure}
      
The correlation between $\rm L_{BOL}/ F_{1.3}^{D_0}$ and N$_2$D$^+$/N$_2$H$^+$ is shown in Fig.~\ref{ll}. The objects with the highest deuterium fractionation ($>0.15$) (IRAS~03282, Barnard~1-B and HH~211) all have low $\rm L_{BOL}/ F_{1.3}^{D_0}$ ratios and are presumably less-evolved protostars. As the $\rm L_{BOL}/ F_{1.3}^{D_0}$ ratio increases, the deuterium fractionation drops quickly. If one considers just the Perseus cores, then the fall-off of the deuterium fractionation is less steep. The $\rm L_{BOL}/ F_{1.3}^{D_0}$ ratios may be lower in the non-Perseus sources because systematic errors in normalisation (see section~\ref{Cod}) lead to overestimates of the continuum fluxes. 
L~1455~A1 has by far the highest $\rm L_{BOL}/ F_{1.3}^{D_0}$ ratio, and by this measure should be the most evolved protostar in the figure. The assertion that L1455~A1 is more evolved that other protostars is reinforced by the relatively low CO depletion factor and high dust temperature. L~1527 is a notable outlier in the sample. Its $\rm L_{BOL}/ F_{1.3}^{D_0}$ ratio is one of the lowest in our sample (5.0), but it has an N$_2$D$^+$/N$_2$H$^+$ ratio $< 0.034$ (more discussion on this source is presented in ~\ref{evo}).

\subsection{Kinematic of the Gas}\label{anav}

So far, we have correlated the N$_2$D$^+$/N$_2$H$^+$ ratios with physical parameters, which are all related to temperature, and thus to the evolutionary stage of the protostar. In this section, we investigate the relation of deuterium fractionation with parameters linked to the kinematics of the gas. Infall, for example, heats up the gas and is expected to influence the gas-phase chemistry.  

First, we compare the N$_2$D$^+$/N$_2$H$^+$ ratios with the N$_2$D$^+$~2-1 line widths (Tab.~\ref{n2h}). Processes such as thermal broadening, turbulent motions, and systematic motions, e.g. infall, contribute all to the full width at half maximum (FWHM) of optically thin emission lines. Thermal broadening plays a relatively minor role, since the thermal line width (FWHM) of  N$_2$D$^+$ at 10~K is only 0.12~km/s (0.17~km/s at 20~K).

\begin{figure}[ht]
   {\resizebox{8cm}{!}
            {\includegraphics{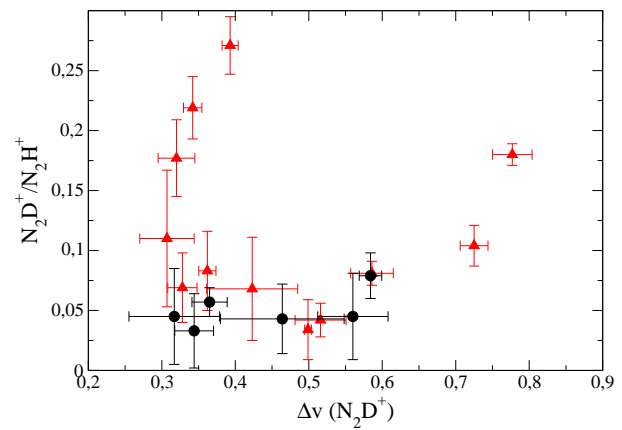}}}    
     \caption{FWHM of N$_2$D$^+$ 2-1 versus N$_2$D$^+$/N$_2$H$^+$. Triangles are Perseus cores.} \label{delv}
      \end{figure}
      
In Fig.~\ref{delv}, the N$_2$D$^+$/N$_2$H$^+$ column density ratio is plotted versus the FWHM of N$_2$D$^+$~(2-1). Most of the protostars with high N$_2$D$^+$/N$_2$H$^+$ ratios have narrow N$_2$D$^+$ lines. The two exceptions are Barnard~1-B ($\rm\Delta v=0.78~km/s$) and L~1448~C ($\rm \Delta v=0.72~km/s$), which show the broadest N$_2$D$^+$ lines in the sample. However, all Perseus sources with $\rm \Delta v>0.5~km/s$, including Barnard~1-B and L~1448~C, have multiple embedded sources (Volgenau~\cite{phd}, Matthews \& Wilson~\cite{b1mult}). The relative velocity differences between the individual protostars leads to additional line broadening.

To be able to trace only the systematic motion of the gas, e.g. infall motion, we used the asymmetry parameter $\rm\delta v$, which was defined by Mardones et al.~(\cite{mard}) as
$$
\rm \delta v= \frac{v_{thick}-v_{thin}}{\Delta v_{thin}}.
$$
v$\rm _{thick}$ is the velocity of an optically thick line, v$\rm _{thin}$ is the velocity of an optically thin line, and $\rm\Delta v_{thin}$ is its line width. As shown by Mardones et al., infall yields a negative value for $\delta$v. If infall dominates the kinematics of the protostar, then an absorbing foregroud component will appear redshifted with respect to the core, causing an optically thick line to appear blueshifted.  
We used HCO$^+$~3-2 as the optically thick line and N$_2$H$^+$~1-0 as the optically thin line. The velocity of the optically thin N$_2$H$^+$ line is derived using the HFS method of CLASS (see section~\ref{res}).  We also compare the resulting velocities with those derived by fitting a single Gaussian to the isolated $1_{01}\rightarrow 1_{12}$ component, following Mardones et al.\cite{mard}. The velocities calculated with the two different methods are not always coincident, but the differences are within the velocity resolution of  0.06~km/s.  Because a fit to seven lines has a lower error than a fit to only one hyperfine component we use in the following the $\rm v_{LSR}$ obtained by the HFS method. 
For v$\rm _{thick}$, we use the peak velocity of the HCO$^+$~3-2 line. In cases where the HCO$^+$ lines are double-peaked, we follow Mardones et al.~(\cite{mard}) and use the velocities of the stronger of the two peaks (although in some spectra the stronger peak was ambiguous). An alternative asymmetry parameter is the ratio of the red peak intensity to the blue peak intensity (R/B) of optically thick, double peaked lines. In such cases, $\rm R/B<1$ indicates infall.
Since more than 50 \% of the observed HCO$^+$ lines show double-peaked line shapes, we also calculate these ratios. The results for both asymmetry parameters are listed in Tab.~\ref{linw}.

\begin{table}[ht]
\caption{Velocities of C$^{18}$O and HCO$^+$ lines and asymmetry parameters. Values for $\delta$v marked with $^*$ are ambiguous due to nearly equal peak intensities of the double-peaked HCO$^+$ line, i.e. the R/B ratio is close to one. } \label{linw}
\begin{center}
\begin{tabular}{lccccc}
\hline
  Source & v [km/s]    &  v [km/s] & $\delta$v  & R/B  \\
  & N$_2$H$^+$  & HCO$^+$  \\
\hline
\hline
L 1448 IRS 2 &  4.12                 & 4.53  & 0.82$^*$ & 1.28\\
L 1448 IRS 3 &  4.62         & 5.40  & 0.84 & - \\
L 1448 C &    4.95       & 5.45  & 0.66 & 2.21 \\
L 1455 A1 &  5.10        & 5.12  & 0.03 & - \\
L 1455 A2 &  4.88        & 4.94  & 0.17 & - \\
Per 4 S &   7.52         & 7.25  & -0.92  & -   \\
Per 4 N &   7.57         & 7.44  & -0.40 & -   \\
Per 5 &   8.15          & 7.65  & -1.31  & -   \\
Per 6 &   5.94           & 6.02  & 0.05 &1.82  \\
IRAS 03282 &  6.06      & 6.62  & -0.91 & 0.57\\
Barnard 1-B & 6.62      & 6.01  & -0.74 & 0.45\\
HH 211 &    9.12         & 9.27  & 0.37$^*$  & 1.47  \\
Barnard 5 IRS 1 & 10.27   & 10.29 & 0.05 & 3.88\\
L 1527 &   5.90            & 5.38  & -1.60 & - \\
L 483 &   5.40           & 5.87  & 1.06  & 3.18\\
SMM 5 &   8.12           & 8.18  & -0.33   & -       \\
L 723 &    11.11          & 11.43 & 0.46   & 1.61      \\
L 673 A &  6.87             & 6.97  & 0.19  & -    \\
B 335 &    8.34         & 7.90  & -1.10$^*$    & 0.833       \\
L 1157 &   2.67         & 3.16  & 0.75  & 1.54      \\
\hline
\end{tabular}
\end{center}
\end{table}

In our sample, we find eight sources with a clear indication of infall (i.e. $\rm\delta v < -0.25$), and seven sources with redshifted, optically thick lines (i.e.  $\rm\delta v > 0.25$). For five sources, there is no clear indication of systematic motion ($-0.25<\delta v<0.25$). Like previous studies by Mardones et al.~(\cite{mard}) and Gregersen et al.~(\cite{greg}), we find more blueshifted than redshifted objects. The blue excess is defined by Mardones et al.~(\cite{mard}) as
$$
\rm blue~excess=\frac{N_{blue}-N_{red}}{N_{total}}
$$
where $\rm N_{blue}$  is the number of objects with $\delta v < -0.25$ and $\rm N_{red}$ is the number of objects with $\delta v > 0.25$. In our sample, the blue excess is $0.05\pm 0.2$. If we exclude the objects with an ambiguous $\delta$v due to nearly equal peak intensities of the double peaked HCO$^+$ line, we derive a blue excess of $0.12\pm 0.2$. Both values are slightly lower, but still agree within the errors with 0.28 and 0.25, the values found by Gregersen et al.~(\cite{greg}) and Mardones et al.~(\cite{mard}),  respectively.

\begin{figure}[ht]
	{\resizebox{8cm}{!}    {\includegraphics{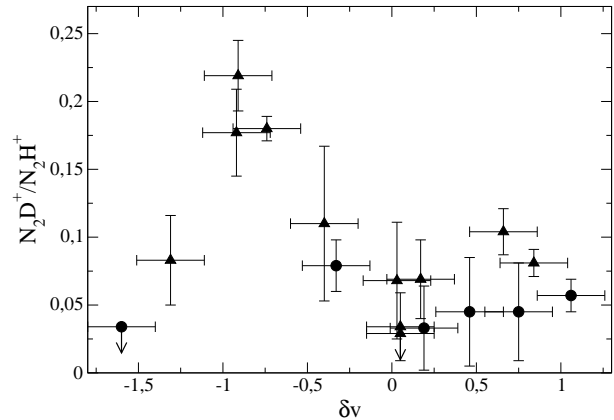}}}    
     \caption{Asymmetry parameter $\delta$v versus N$_2$D$^+$/N$_2$H$^+$ ratio. Only the sources with an unambiguously determined asymmetry parameter are shown.     
     All sources with enhanced deuterium fractionation are located at moderate $ \delta v$ of about -0.4, and is increasing with decreasing deuterium fractionation.} \label{dv}
      \end{figure}

$\delta v$ and, less clearly, R/B are correlated with the deuterium fractionation, and only Per~5 and, once again L~1527, which is a peculiar source (see section~\ref{evo}), do not follow the general trend and show indications of strong infall motion. The higher the N$_2$D$^+$/N$_2$H$^+$ ratio is, the more blueshifted the optically thick emission lines are. Such a result was somehow expected, since outflow activity rises with the evolution of the protostar (Richer et al.~\cite{pp5}). The results are also in good agreement with the results of Mardones et al.~(\cite{mard}). Using observations of CS, C$^{34}$S and H$_2$CO lines, they found that the blue excess of Class~0 protostars is clearly enhanced compared to Class~I objects.
However, using observations of HCO$^+$, H$^{13}$CO$^+$ and N$_2$H$^+$ lines, Gregersen et al.~(\cite{greg}) did not find such trend. They speculated that infall asymmetry disappears later in HCO$^+$. Because our sample contains no true Class~I objects, we cannot directly compare their results with ours.

We also calculated $\delta v$ using the C$^{18}$O~1-0 transition as the optically thin line. Contrary to N$_2$H$^+$, which traces the cold, CO depleted envelope, C$^{18}$O traces the already warmed up central part of the protostar. Thus it traces the region, where the HCO$^+$~3-2 line stems from, too. But due to the lower critical density of the C$^{18}$O~1-0 transition, the low density outskirts of the protostellar envelope might contribute significantly to the C$^{18}$O emission, which possibly dilute any correlation. Qualitatively, the correlation between $\delta v$ and the deuterium fractionation looks similar, no matter which optically thin line is used.  However, for objects with an N$_2$D$^+$/N$_2$H$^+$ ratio above 0.75, $\delta v$ determined with C$^{18}$O yields to less blue-shifted values. A reason for this difference might be that in these objects, which also show the highest CO depletion factors, the contributions from the environmental material to the C$^{18}$O emission is expected to be biggest.  The bigger contribution of the environmental material is also indicated by the fact that for objects with $\delta v_{N_2H^+}$ lower than -0.5, $\rm\Delta v(C^{18}O)$ is on average 2.4 times larger than  $\rm\Delta v(N_2H^+)$, whereas for the remaining sources $\rm\Delta v(C^{18}O)$ is only 1.6 times larger. More turbulent material appears to surround infalling protostellar envelopes.

%For all asymmetry parameters ($\delta$v, R/B, and $\rm |\delta v|$), no clear correlation with deuterium fractionation could be found (see Fig~\ref{dv} for $\rm| \delta v|$ versus N$_2$D$^+$/N$_2$H$^+$). However, sources with high deuterium fractionation tend to show infall motions. Three out of the four sources with an N$_2$D$^+$/N$_2$H$^+$ ratios above 0.15 have negative $\delta v$ and the infall parameter of the fourth object (HH~211) is ambiguous, because of double peaked spectra with almost equally high peak intensities. But the other objects show no indication of a correlation between $\delta v$ and deuterium fractionation. Hence the kinematics of the gas plays no or only a minor role in deuterium chemistry.  
     
\subsection{Comparison of the deuterium fractionation of different molecules}\label{anam}

The deuterium fractionation of N$_2$H$^+$ reflects the present time temperature and density better than most other molecules, since it does not significantly deplete (see e.g. Caselli et al.~\cite{cas2}, Bergin et al.~\cite{ted}, Schnee et al.~\cite{scg07}) and the chemical reaction for N$_2$D$^+$ formation and destruction are quite fast and simple. Memories of  cold phases in the past of the dark cloud core are therefore not expected to be seen in the observed  N$_2$D$^+$/N$_2$H$^+$ ratio. In this subsection, we compare the N$_2$D$^+$/N$_2$H$^+$ ratio with the NH$_2$D/NH$_3$ (Hatchell~\cite{nh3}), DCO$^+$/HCO$^+$ ratios (J\o rgensen et al.~\cite{jor}) and the deuterium fractionation of H$_2$CO and CH$_3$OH (Parise et al.~\cite{d2co}) for the sources in our sample where these data are available (Tab.~\ref{D/Ho}).

\begin{table}[ht]
\caption{Deuterium fractionation of several molecular species.} \label{D/Ho}
\begin{center}
\begin{tabular}{lcccc}
\hline
%Source & N$_2$D$^+$/N$_2$H$^+~^1$ & NH$_2$D/NH$_3~^2$ & DCO$^+$/HCO$^+~^3$ \\
Source & $\rm\frac{N_2D^+}{N_2H^+}~^1$ & $\rm\frac{NH_2D}{NH_3}~^2$ & $\rm\frac{DCO^+}{HCO^+}~^3$ & $\rm\frac{D_2CO}{H_2CO}~^4$ \\
\hline
\hline
L1448~IRS~3 & 0.081 & 0.17 & - & -\\ 
L1448~C & 0.104 & 0.20 & 0.011 & 0.24\\
IRAS~03282 & 0.219 & 0.22 & - & -\\
HH 211 & 0.271 & 0.33 & - & -\\
Barnard 5 & 0.034 & 0.18 & - & -\\
L 1527 & $<$ 0.034 & - & 0.048 & 0.44 \\
L 483 & 0.057 & - & 0.006 & -\\
L 723 & 0.045 & - & 0.004 & -\\
L 1157 & 0.045 & - & 0.016 & $\leq 0.08$\\  
\hline
\end{tabular}
\end{center}
$^1$This work, $^2$Hatchell~(\cite{nh3}), $^3$J\o rgensen et al.~(\cite{jor}), $^4$ Parise et al.~(\cite{d2co}).
\end{table}

The deuterium fractionation of HCO$^+$ is, in all objects but one (L~1527), significantly lower than the fractionation of N$_2$H$^+$. 
This difference can be explained by the fact that CO, the parent species of HCO$^+$, is also the main destroyer of H$_2$D$^+$ and thus the DCO$^+$/HCO$^+$ ratio traces only regions where the deuterium fractionation in N$_2$H$^+$ is expected to be lower.
Furthermore, J\o rgensen et al.~(\cite{jor}) determined the DCO$^+$/HCO$^+$ ratio using the DCO$^+$~J=3-2 line, which has a critical density of $\rm 1.4\cdot 10^6~cm^{-3}$.  Therefore this DCO$^+$ observations traces mostly the inner, high density part, which might be already warmed up by the embedded protostellar core, thus lowering down the DCO$^+$/HCO$^+$ ratio.
The situation with NH$_3$ is different. Like N$_2$H$^+$, NH$_3$ is formed from N$_2$ and does not seem to freeze out (Tafalla et al~\cite{taf}). Thus, N$_2$D$^+$ and NH$_2$D emission should originate from the same region. As a consequence, the deuterium fractionation of N$_2$H$^+$ and NH$_3$ should be comparable.

\begin{figure}[ht]
	{\resizebox{8cm}{!}    {\includegraphics{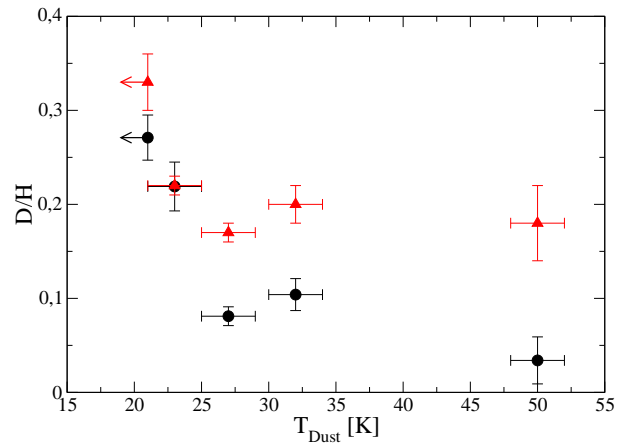}}}    
     \caption{Deuterium fractionation of NH$_3$ (red triangles) and N$_2$H$^+$ (black circles) versus dust temperature. } \label{figNH3}
      \end{figure}
      
As shown in Fig.~\ref{figNH3},  the coldest objects, i.e. the youngest objects in our sample (HH~211 \& IRAS~03282), have a deuterium fractionation of NH$_3$ indeed comparable to the fractionation of N$_2$H$^+$. But the warmer the dust gets, the more the ratios diverge. The NH$_2$D/NH$_3$ ratio is about 0.2 at  T$\rm _{Dust}$ = 50~K, where the N$_2$D$^+$/ N$_2$H$^+$ ratio is five times lower. 
The differing trends in the two ratios can be explained as follows. Both N$_2$D$^+$ and NH$_2$D are formed directly or indirectly via H$_2$D$^+$ or its multiply deuterated forms, i.e. the formation mechanism is only effective in cold, CO-depleted environments. In regions were CO desorbs from the dust due to the warming up of the protostar, N$_2$D$^+$ and N$_2$H$^+$ disappear quickly from the gas phase, because these species are directly destroyed by CO, forming HCO$^+$ and DCO$^+$, respectively. Thus, with increasing temperature, the N$_2$D$^+$/N$_2$H$^+$ ratio traces material at larger and lager radii (i.e. at lower and lower densities).  But deuterium fractionation is related to the density, because of the lower degree of CO depletion at lower densities (see Sect.~\ref{strat} and Fig.~\ref{COHD}).  Contrary to N$_2$H$^+$ and N$_2$D$^+$, which are pure gas-phase species, ammonia is also found on dust grain mantles, most likely formed by grain-surface reactions (Boogert et al.~\cite{eis}). The solid NH$_3$, which desorbs form the dust grains as soon as the environmental material warms up, likely has a large deuterium fractionation due to the fossil record of grain surface chemistry and gas phase depletion at low temperatures (e.g. Aikawa et al.~\cite{awg08}). Thus, the observed NH$_2$D/NH$_3$ ratio declines more slowly than the deuterium fractionation of pure gas phase species.  

Deuterated formaldehyde (HDCO and D$_2$CO) are very abundant, too. The HDCO/H$_2$CO ratio as well as the D$_2$CO/H$_2$CO ratio are of the order of 0.1 (Parise et al.~\cite{d2co}). In their work, Parise et al. also determined the deuterium fractionation of methanol, which exceeds even the deuterium fractionation of formaldehyde (e.g. $\rm CH_2DOH/CH_3OH\sim 0.5$). For both molecules, no obvious correlation between the deuterium fractionation and the core evolutionary stage could be found. Because mainly grain surface chemistry is involved in the deuteration of methanol and formaldehyde, the observed deuterium fractionation is likely a memory of the previous, prestellar phase (Ceccarelli et al.~\cite{clc01}), similar (although more extreme) case as found for ammonia. 

\subsection{Evolutionary sequence of the sample}\label{evo}

%The dust temperature, CO depletion, and $\rm L_{BOL}/F^{220 pc}_{1.3}$ all correlate with the degree to which a protostar has evolved since the moment of star formation. Here, we compare these indicators of protostellar evolution to the deuterium fractionation to define an evolutionary sequence among the sources in our sample . 

The youngest objects in our sample are Barnard~1-B, IRAS~03282 and HH~211. All three evolutionary tracers indicate those sources to be among the youngest. The depletion factor, f$\rm _D$(CO), indicates that L1448~IRS3 and L~1157 are young objects, too, but their T$\rm_{DUST}$ suggest that these sources are, instead, more evolved. A possible solution to this discrepancy lies in acknowledging several assumptions we made in determining the CO depletion factor (e.g. in converting the beam sizes of the continuum and the C$^{18}$O observations). Although, overall, these assumptions seem to be justified, they might not be valid for all sources and thus the f$\rm _D$ (CO) values may be less accurate. The three protostars we identified as the youngest also have the highest N$_2$D$^+$/N$_2$H$^+$ ratios (0.18, 0.22 and 0.27 in Barnard~1-B, IRAS 03282 and HH~211, respectively). By this measure, Per~4~S, with a deuterium fractionation of 0.177, should also be very young, but no complementary data are available to confirm this. 

The objects at a moderate evolutionary stage are those with a deuterium fractionation around 0.1 (L~1448~IRS~3, PER~4~N, PER~5). The exception  is L~1448~C (N$_2$D$^+$/N$_2$H$^+ = 0.10$), which seems to be more evolved than the others. However, among these intermediately-evolved sources, there are also some with deuterium fractionations of $\sim 0.05$ (L~723, B~335, L~1448~IRS~2). For SMM~5 (N$_2$D$^+$/N$_2$H$^+ = 0.079$), we can't determine the evolutionary stage independently, but, because of its N$_2$D$^+$/N$_2$H$^+$ ratio, we think that it belongs to the group of intermediately-evolved objects. The N$_2$D$^+$/N$_2$H$^+$ ratios of the most evolved protostars are, with the exception   of L~1448~C (0.10), $\sim 0.05$.

L~1527 is a special case. The T$\rm _{Dust}$ (27~K) and especially the $\rm L_{BOL}/ F_{1.3}^{D_0}$ ratio (5.0) suggest that it is a younger object, but its low N$_2$D$^+$/N$_2$H$^+$ ratio and the fact that an embedded source is detected at wavelengths $< 5\mu m$  indicates that this source is more evolved (Froebrich~\cite{fro}). Chemical models, which reproduce the observed abundance of long-chain unsaturated hydrocarbons and cyanopolyynes, indicate a more evolved evolutionary stage of L~1527 as well (Hassel et a.~\cite{has}). From infra-red observations, Tobin et al.~(\cite{1527}) concluded that L~1527 contains a massive protostellar disk, which is seen edge-on. Therefore  L~1527 might be a more evolved object than indicated by continuum observations, due to the large obscuration of the central object. Hence, the N$_2$D$^+$/N$_2$H$^+$ ratio seems to be even more reliable than continuum measurements to identify the youngest protostellar objects. 

In conclusion, the N$_2$D$^+$/N$_2$H$^+$ ratio can be used to clearly indentify the youngest objects among the Class~0 sources. For more evolved objects, the correlation of evolutionary stage and  N$_2$D$^+$/N$_2$H$^+$ ratio is not so clear, but an overall decrease of deuteration with evolutionary stage can be seen.

\section{Comparison with chemical- and radiative transfer models}\label{model} 

To explain the results discussed in the previous section, we examined a model of a ``typical''  Class~0 object rather than a specific source. First, we describe the physical properties, e.g. the temperature and density profiles, of the model. Subsequently, we calculate the abundances of the important species and apply a radiative transfer code to simulate the observations.

\subsection{Cloud structure}

 The density structure of the model is that of a singular isothermal sphere (Shu~\cite{s77}):
$$
\rm n(r)\propto r^{-2}.
$$
Such density profiles have been found for Class~0 protostars, too (Lonney et al.~\cite{env}).
The inner boundary of the protostellar envelope is set to 100~AU. This limit is arbitrarily chosen, but at a distance of 220~pc, which is the distance of the Perseus molecular cloud, 100~AU corresponds to an angle of $ 0.46''$, which is much smaller than the angular resolution of our observations. Therefore, the model results are not sensitive to the chosen inner boundary. The outer boundary is defined by the radius where the density is $\rm 10^4~cm^{-3}$. Kirk et al.~(\cite{pre}) estimated the external pressure on cores in the Perseus region and found values of $\rm log_{10}(P_{ext}/k)$ between 5.5 and 6.0. Assuming a temperature between 15 and 30~K, this leads to a gas density between $\rm\sim 10^4~cm^{-3}$ and $\rm\sim 5\cdot 10^4~cm^{-3}$. Around L~1448~IRS~3, Kwon et al.~(\cite{dens1448}) find a density of $\rm 6-9\cdot 10^3~cm^{-3}$. Thus, $\rm 10^4~cm^{-3}$ can be considered an average value for the gas number density surrounding Perseus cores.

\begin{figure}[ht]
	{\resizebox{9cm}{!}    {\includegraphics{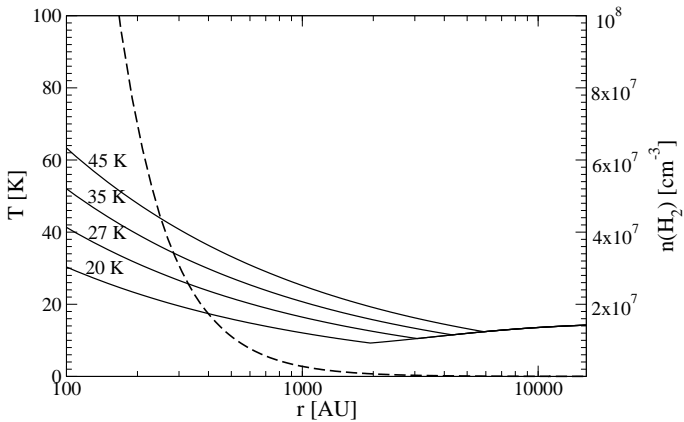}}}    
     \caption{Density (dashed line) and temperature (solid lines) profiles of the protostellar core models. The four temperature profiles represent models that yield values for T$\rm _{Dust}$ of 20~K, 27~K, 35~K and 45~K, respectively.} \label{profiles}
      \end{figure}

The total envelope mass is assumed to be of the order of one to a few solar masses. However, since the model results depend strongly on the density, the total mass varies for different models (see below).
For the temperature profile, we use a power law with an index of -0.4, which is the exponent for optically thin emission from protostellar envelopes, as found by Looney et al.~(\cite{env}). They also showed that, with a density power-law index of 2, the radial temperature distribution can significantly deviate from the power-law at radii smaller than 100-200 AU. However, given that we used an inner boundary of 100~AU for the envelope, such deviations do not noticeably influence our model.
To avoid very low temperatures in the outskirts of the envelope, we assume a temperature profile like that found in prestellar cores (Crapsi et al.~\cite{temppre}):
$$
\rm T(r)=T_{out}-\frac{T_{out}-T_{in}}{1+(r/r_0)^{1.5}}.
$$  
Here, we assume T$\rm _{out}=15$~K, T$\rm _{in}$=7~K, and r$_0$=3633~AU. This radius corresponds to an angle of $26''$ at a distance of 220~pc, as found in L~1544. However, since we use this temperature profile only for the outer, low-density part of the envelope, the influence of these parameters on the model results is negligible.
Density and temperature profiles are shown in Fig.~\ref{profiles}.

Using this cloud structure, we calculate the continuum flux at 60~$\rm \mu m$, 100~$\rm \mu m$, 850~$\rm \mu m$ and 1.2~mm from the formula:
$$
\rm F_\nu=\int n(r)\cdot B_\nu(T(r))\cdot \kappa_\nu\cdot\mu~dr
$$
where n is the density, B(T) is the Planck function, and $\mu$ is the mass of the gas molecules. For the opacity per unit mass, $\kappa _\nu$, we use again 0.005 g/cm$\rm^{2}$ at 1.2 mm, which assumes optically thin continuum emission. Furthermore, we assumed an emissivity index of $\beta=2$, which is a typical value for molecular clouds (Ward-Thompson et al.~\cite{war}) and not the $\beta = 1$ shown by previous fits. We used $\beta =2$, because the cold ($\sim 10$~K) outskirts of the envelope still emit a significant amount of radiation at mm wavelengths, but much less at 60~$\rm \mu m$. The strong temperature gradient in protostellar cores causes the spectrum integrated over the entire core to be flattened. Thus, the emissivity index derived from a single temperature fit is underestimated.
The values for the continuum emission obtained by this method are then fitted, following the procedure described in section~\ref{Tdustsec}, and the absolute temperature scale is varied to obtain models with dust temperatures of 20~K, 27~K, 35~K and 45~K, respectively, which covers the whole range of observed values.

\subsection{Chemical and radiative transfer model}\label{codes}

The chemical code used in this study is similar to the one originally described by Caselli et al.~\cite{cas2}, but updated following Caselli et al.~\cite{cvv08}.  We assume a spherically symmetric cloud, and we follow the time-dependent freeze-out of CO and N$_2$, as well as thermal and non--thermal desorption due to cosmic--ray impulsive heating, neglecting dynamical effects. All the deuterated forms of H$_3^+$ are included in the model, and the abundances of the ionic species such as HCO$^+$, N$_2$H$^+$ and their deuterated couterparts are given by the steady state chemical equations using the instantaneous abundaces of the neutral species (see Caselli et al.~\cite{cas2} for details).  The rate coefficients have been adopted from the UMIST database for astrochemistry, available at http://www.udfa.net. For the proton--deuteron exchange reaction (such as H$_3^+$ + HD 
$\rightarrow$ H$_2$D$^+$ + H$_2$), we used both ``standard rates'' (e.g. as used in Roberts et al.~\cite{rhm04}, Ceccarelli \& Dominik~\cite{cd05}) as well as the rates more recently measured by Gerlich et al.~(\cite{ghr02}), which appear to better fit our data (see below). No atomic oxygen is included in the code, given the large uncertainties in its abundance in cold and dense regions (e.g. Bergin \& Snell~\cite{bs02}, see also discussion in section 3.2.3 of Caselli et al.~\cite{cas2}). The cosmic-ray ionization rate has been fixed at $\zeta$=3$\times$10$^{-17}$~s$^{-1}$, the average value measured toward high-mass star forming regions by van der Tak \& van Dishoeck~(\cite{vv00}).  The CO and N$_2$ binding energies are assumed to be 1100~K and 982.3~K, respectively, following the prescriptions of \"{O}berg et al.~\cite{ovf05} and Schnee et al.~\cite{scg07} (see their section 6.2).  The model is run for 10$^6$~yr, an average age for a dense core. After  10$^6$ years, the abundances of the various molecules have nearly reached chemical equilibrium, and thus the exact evolution time is not a critical parameter (see section~\ref{sensi} for discussion).

The results from the chemical model are used as input parameters for a radiative transfer model,  a non-LTE Monte-Carlo code developed by Pagani et al.~(\cite{MCn}). This code, which is developed from the Bernes code (Bernes~\cite{MCo}), calculates the 1D radiative transfer of both N$_2$H$^+$ and N$_2$D$^+$ and includes line overlap of the hyperfine transitions. Furthermore, recently published collisional coefficients between individual hyperfine levels have been implement (Daniel et al.~\cite{coll}). For the convolution with the beam, we assumed a Gaussian beam shape with HPBW as listed in Tab.~\ref{lines}. The distance to the cloud is set to 220~pc, i.e. the distance of the Perseus cloud. From the spectra that we obtained from the Monte-Carlo-code, we determined the N$_2$D$^+$/N$_2$H$^+$ ratio using the CTEX procedure, just as we did for the observed spectra (see section~\ref{ana1}). Hence, the comparison between the model and the observations is consistent, because possible systematic errors, introduced by the assumption of a constant excitation temperature, are taken into account.

\subsection{Comparison between models and observations}

\begin{figure}[ht]
	{\resizebox{8cm}{!}    {\includegraphics{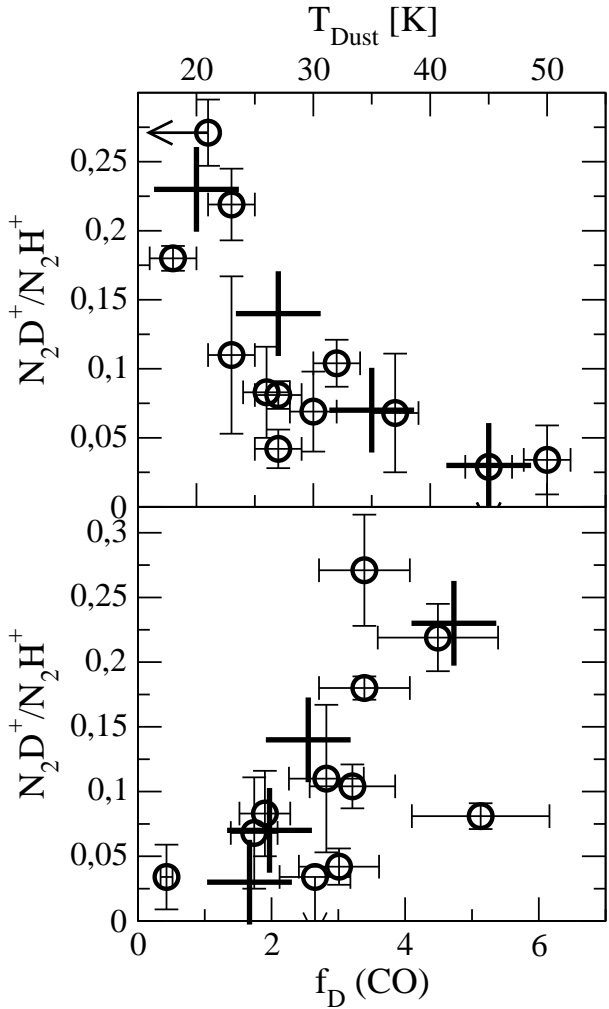}}}    
     \caption{{\bf{Upper panel:}}N$_2$D$^+$/N$_2$H$^+$ ratio versus dust temperature. {\bf{Lower panel:}} N$_2$D$^+$/N$_2$H$^+$ ratio and the CO depletion factor F$\rm_D$~(CO). 
     The open circles are the observed values for the sub-sample of cores in the Perseus cloud; the crosses are the model results. The N$_2$D$^+$/N$_2$H$^+$ ratio of the models was determined using the CTEX method.} \label{Tdustmodel}
      \end{figure}

We compute the N$_2$D$^+$/N$_2$H$^+$ ratios for four models with fitted dust temperatures of 20~K, 27~K, 35~K and 45~K, respectively. The deuterium fractionation in these models ranges from 0.23 at 20~K to 0.03 at 45~K. These values fit the observations very well. We derive an H$_2$ density at a radius of 100 AU of  $\rm 2.77\cdot 10^8~cm^{-3}$, which corresponds to a total envelope mass of 4.6~$\rm M_{\odot}$. 
The modelled correlation between the deuterium fractionation and the CO depletion factor (f$\rm_D$~(CO)) fits the observations quite well, too, but in the cooler objects (T$\rm _{Dust} \sim 20$~K), the modelled CO freeze-out is very sensitive to the temperature. Thus, a marginal temperature variation leads to a significant change of the CO depletion factor, and therefore the quality of the fit depends strongly on the model used. 
In Fig.~\ref{Tdustmodel}, a comparison of the models with the observations of the Perseus sources is shown. Results from the models are also listed in Table~\ref{modres}.
The best agreement with observations is found when using the reaction rates measured by Gerlich et al.~\cite{ghr02} for the proton-deuteron exchange reaction. Using the ''standard rates'', we could only fit the observed N$_2$D$^+$/N$_2$H$^+$ ratios by increasing the total envelope mass to 25~M$_\odot$, a value that is clearly larger than expected for low mass protostars.

To check that the assumption of a constant excitation temperature does not give a significantly different result from the $\rm T^{-0.4}$ profile, we calculated beam averaged column densities from the model abundance profiles as well. The N$_2$D$^+$/N$_2$H$^+$ ratios determined from the beam averaged column densities are the same as before, to within 20\%. The estimated error of the observed N$_2$D$^+$/N$_2$H$^+$ ratio is also approximately $20\%$, so the fit is good, independently of the method used. 
The trend of a decreasing deuterium fraction with increasing T$\rm _{Dust}$ is true for both the ratio determined by the CTEX method and the ratio of the beam averaged column density. 

\begin{table}[ht]
\caption{Model results} \label{modres}
\begin{center}
\begin{tabular}{cccc}
\hline
T$\rm_{Dust}$ & N$_2$D$^+$/N$_2$H$^+$ (CTEX)  & HCO$^+$/DCO$^+$ & f$\rm _D$(CO)\\ 
\hline
\hline
20 K & 0.23 & 0.09 & 4.73 \\
27 K & 0.14 & 0.05 & 2.55 \\
35 K & 0.07 & 0.03 & 1.97 \\
45 K & 0.03 & 0.02 & 1.67 \\
\hline
\end{tabular}
\end{center}
\end{table} 

The N$_2$D$^+$ J=3-2/J=2-1 intensity ratios of the models range from 0.44 to 1.10, which is comparable to the observed ratios. For the  J=2-1/J=1-0 ratios, the models give values a factor of two higher than what is observed. However, these ratios are very sensitive to temperature, density and abundance variations. Taking the simplicity of our model into account, we consider these values in agreement with the observations.  

Besides N$_2$D$^+$ and N$_2$H$^+$, the abundances of HCO$^+$ and DCO$^+$ are also calculated by the chemical model. The DCO$^+$/HCO$^+$ ratio is significantly lower than the N$_2$D$^+$/N$_2$H$^+$ ratio and ranges from 0.02 to 0.08 in the models with dust temperatures of 45~K and 21~K, respectively. The HCO$^+$ and DCO$^+$ column densities are averaged over a 21$''$ beam, and thus comparable to the observations conducted by J\o rgensen et al.~(\cite{jor}). The values for the DCO$^+$/HCO$^+$ ratio reported by them are between 0.004 and 0.048 and thus a little lower than the modelled ratios (see Tab.~\ref{modres} \& \ref{D/Ho}).

\subsection{Chemical stratification within the core}\label{strat}

\begin{figure}[ht]
	{\resizebox{8cm}{!}    {\includegraphics{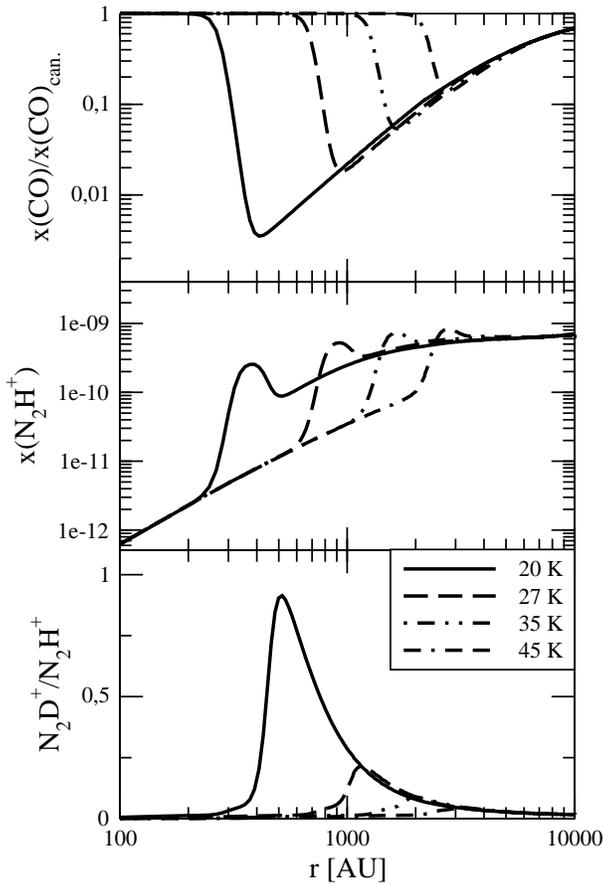}}}    
     \caption{The relative abundance of CO (upper panel), N$_2$H$^+$ (middle panel) and the N$_2$D$^+$/N$_2$H$^+$ ratio are shown as a function of radius. It can be clearly seen that the radius at which CO starts to deplete, as well as the peak of the deuterium fractionation, shifts outwards with increasing temperature.} \label{COHD}
      \end{figure}

The temperature gradient, caused by the heating of the central object, leads to strong variations of the abundances of many molecules and molecular ions. The more immediate cause for these variations is the freeze-out of CO, which occurs at a temperature of $\sim 20$~K. The absence of CO in the gas phase has a big influence on the abundance of many other species, since many molecular ions, e.g. N$_2$H$^+$, are partially destroyed by reactions with CO. In addition, the deuterium fractionation is enhanced in regions where CO is depleted (see section~\ref{intr}). Therefore, the protostellar envelope can be roughly divided into three zones (Fig.~\ref{COHD}). In the inner zone, the temperature has increased above $\sim 20$~K, CO  evaporates from the grains and, consequently, the abundance of N$_2$H$^+$ as well as the deuterium fractionation are relatively low in this region. As soon as the temperature drops below the critical value of $\sim 20$~K, the CO abundance drops drastically. At the same time, the N$_2$H$^+$ abundance rises by 1-2 orders of magnitude. The N$_2$D$^+$/N$_2$H$^+$ ratio rises as well, but its maximum is at a larger radius than the CO abundance minimum. This is because the temperature at the minimum of the CO abundance is $\sim 20$~K, where the destruction of H$_2$D$^+$ via the reaction $\rm H_2D^+ + H_2 \rightarrow H_3^+ + HD$ is not neglible.
At even larger radii, the degree of CO depletion decreases as the gas density decreases, which causes the N$_2$H$^+$ abundance to stay more or less constant, and the deuterium fractionation to drop again.
 
During the Class~0 phase, the temperature of the central object increases, and so does the temperature of the envelope. Consequently, the radius at which CO freezes out onto dust grains shifts further outwards, and the fraction of the cloud in which the abundance of deuterated molecules is enhanced gets smaller with time. Furthermore, the mean density of the deuterium fractionated gas decreases, thus the mean local N$_2$D$^+$/N$_2$H$^+$ ratio declines as well. This leads to the decrease of the  N$_2$D$^+$/N$_2$H$^+$ column density ratio, which we observe in our sample. 
      
\begin{figure}[ht]
	{\resizebox{8cm}{!}    {\includegraphics{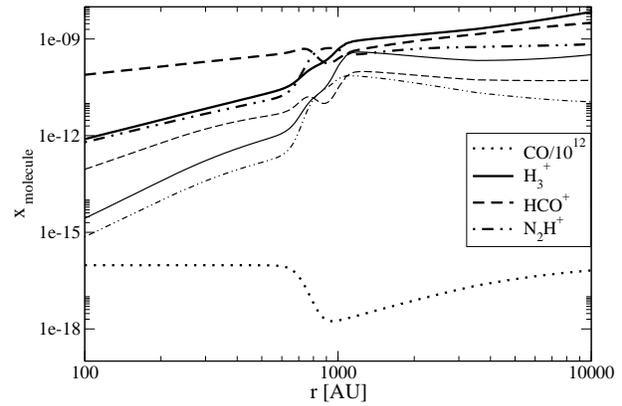}}}    
     \caption{The abundances of CO, H$_3^+$, H$_2$D$^+$, HCO$^+$, DCO$^+$, N$_2$H$^+$, and N$_2$D$^+$ relative to H$_2$ are shown as a function of radius. The thick lines show the abundance profile of the protonated  species. The thin lines show their singly-deuterated counterparts. These profiles apply to the model with T$\rm _{Dust}=27~K$.} \label{molec}
      \end{figure}

The reason for the different ratios of the deuterated and non-deuterated forms of  HCO$^+$ and N$_2$H$^+$ is that the deuterium fractionation is strongest in places where CO, a progenitor of HCO$^+$ \& DCO$^+$, is mostly depleted. Thus, contrary to N$_2$H$^+$, HCO$^+$ is also relatively strong in the warm and dense innner part of the envelope, and the ratio of the column densities of DCO$^+$ and HCO$^+$ is significantly lower than the N$_2$D$^+$/N$_2$H$^+$ ratio. This can be seen in Fig.~\ref{molec}, in which the abundance of the molecules and molecular ions are shown as a function of radius.

The radial abundance profiles we found with our static model are in good agreement with the results of Lee et al.~(\cite{lee}), who calculated a self-consistent dynamical and chemical model of the entire star forming process. In their model, the deuterated species are not calculated, but the abundance profiles of CO, N$_2$H$^+$ and HCO$^+$ at a time $\sim 10^5$~yr after the collapse starts are similar to the profiles we found (see their Fig.~11). This similarity indicates that the assumption of a chemical equilibrium is valid at least for the simple species.  
In a more recent paper, Aikawa et al.~(\cite{aik}) get somewhat shorter timescales for the protostellar evolution:  $9.3\cdot 10^4$ years after the beginning of the collapse, the CO desorption radius is $\sim 2000$~AU, which is comparable to the results of  our 45~K model. But even for these shorter time scales, the assumption of chemical equilibrium is still valid (see next section).

\subsection{Sensitivity of the Model}\label{sensi}

%In this section, we discuss how variations of the parameters which were fixed in our model, such as the cosmic-ray ionization rate and binding energies of CO on dust, influence the resultant N$_2$D$^+$/N$_2$H$^+$ ratio. 

The evolution times of the models are the least constrained values. The estimated ages of Class~0 protostars range from $\sim 10^4$ to a few times $10^5$ years after the collapse has started (Froebrich~\cite{fro}). The ages vary  by a factor of 10 depending which model is used.  However, the crucial time for the deuterium chemistry is the time since the CO started to freeze out, because the  freeze-out of CO is the slowest reaction in the network. Thus, the entire prestellar phase has to be taken into account.  After a time of $10^6$ years, the abundances have already reached equilibrium. Reducing the time to 10$^5$ years changes  the N$_2$D$^+$/N$_2$H$^+$ ratio only marginally. The difference between the ratios after $10^6$ years and $10^5$ years is less than 0.01. In models with a evolution time of 10$^4$ years, the abundances are no longer in equilibrium. In the coldest protostar (20~K), the deuterium fractionation is already high (0.26), but for the warmer objects, the N$_2$D$^+$/N$_2$H$^+$ ratio is a factor $\sim 2$ lower than in chemical equilibrium. Although already quite low, these values still agree with the observations. Evolution times longer than $10^7$ years do not change the results, because equilibrium is already reached. 

Another important parameter is the cosmic-ray ionisation rate. In our model, we used $\rm \zeta =3\cdot 10^{17}~s^{-1}$ (van der Tak \& van Dishoeck~\cite{vv00}). A $\zeta$ of $\rm 6\cdot 10^{17}~s^{-1}$ increases the N$_2$H$^+$ abundance by a factor of 1.5, but the deuterium fractionation is lowered by 40\%. A cosmic-ray ionization rate of $\rm 1.5\cdot 10^{17}~s^{-1}$ increases the N$_2$D$^+$/N$_2$H$^+$ ratio by 30\%.  

%%% NHV: What is the uncertainty in zeta? How well is it constrained?

The central density of the core, and consequently the envelope mass, has a similar effect on the deuterium fractionation as $\zeta$. In our models, the density is varied to match the observed N$_2$D$^+$/N$_2$H$^+$ ratio best. The only constraint we put on the central density is that the resulting envelope mass should be of the order of a few solar masses. For our best fit model, we derive an envelope mass of 4.6~M$_\odot$. Decreasing the density by a factor of a half, the deuterium fractionation decreases by 40\%. With a density twice as high as our best fit model, the  N$_2$D$^+$/N$_2$H$^+$ increases by a factor of 1.2. 

The last crucial parameter we investigated is the binding energy of CO and N$_2$ on dust grains. In our model, we used binding energies of 1100~K and 982.3~K for CO and N$_2$, respectively, which are the binding energies on mixed CO-H$_2$O ice. We also calculated the models using the values for $\rm E_b$ on pure CO ice (855 and 790 for CO and N$_2$, respectively, {\"O}berg et al.~\cite{ovf05}). With these values, the resulting N$_2$D$^+$/N$_2$H$^+$ ratios drop to $<0.04$ for all four temperatures.  

\section{Summary}

We observed 20 Class~0 protostars in N$_2$H$^+$~1-0, N$_2$D$^+$~1-0, 2-1 and 3-2, C$^{18}$O~1-0 and HCO$^+$~3-2. The integrated intensities of the N$_2$H$^+$~1-0 lines in  our sample are  comparable to those found in prestellar cores (Crapsi et al.~\cite{crap}). The optical depths are typically 67\% lower in the Class~0 sources, but because the excitation temperature is 2-3~K higher and the line width is about 2.5 time lager, the N$_2$H$^+$ column densities ($\rm \sim 10^{13}~cm^{-2}$) are also comparable. The presence of a protostar also affects the kinematics of the cores. N$_2$H$^+$ lines are significantly wider in Class~0 sources than in prestellar cores (on average, 0.61~km/s and 0.26~km/s, respectively). A comparison of our HCO$^+$ observations with observations conducted with a $\sim$three times larger beam (Gregersen et al.~\cite{greg}) leads us to the conclusion that HCO$^+$~3-2 stems  mainly from a compact region of about 4000~AU in size.         

The dependence of the N$_2$D$^+$/N$_2$H$^+$ ratio on dust temperature, CO depletion factor, and the L$\rm _{BOL}$/F$\rm ^{220~pc}_{1.3}$ ratio clearly indicates a close relation between deuterium fractionation and evolutionary stage of a Class 0 protostar. For the sub-sample of sources in the Perseus cloud, the correlation is striking. This might be an indication for the influence of the environment on the deuterium fractionation.

The correlation of the deuterium fractionation and the CO depletion factor looks qualitatively like the one found in prestellar cores (Crapsi et al.~\cite{crap}). However, whereas in prestellar cores, the N$_2$D$^+$/N$_2$H$^+$ ratio stays low until f$\rm _{D}(CO)\sim 10$, in protostars, the ratio starts to rise at f$\rm _{D}(CO)\sim 3$. This difference occurs because CO is highly depleted at the centre of prestellar cores, but not at the center of protostellar cores.

There is a weak correlation of N$_2$D$^+$ line width with N$_2$D$^+$/N$_2$H$^+$ ratio, but line broadening by multiple sources makes this conclusion suspect. A clear correlation with other kinematical tracers, especially with $\delta$v, is found. Sources with a high N$_2$D$^+$/N$_2$H$^+$ ratio clearly show infall motion ($\rm\delta v\sim -0.4$). The lower the deuterium fractionation gets, the more $\delta$v increases. This might reflect the fact that outflow activity increases with evolutionary stage (Richer et al.~\cite{pp5}).

%%% NHV: Last sentence: I thought that outflow activity DECREASES with evolutionary stage. Class 0 YSOs have stronger outflows than Class I YSOs.

A model with a power law density and temperature profile reproduces the correlation between the N$_2$D$^+$/N$_2$H$^+$ ratio and both T$\rm _{Dust}$ and the CO depletion factor very well. 
In these models, the protostellar envelopes are chemically stratified. The inner part is too warm for CO to freeze out, so that the N$_2$H$^+$ abundance and the deuterium fractionation are low. At radii where the temperature drops below $\sim 20$~K, CO is mostly depleted and the N$_2$D$^+$/N$_2$H$^+$ ratio increases by several orders of magnitude. At even larger radii, the lower density decreases the degree of CO depletion, and thus the N$_2$D$^+$/N$_2$H$^+$ ratio also decreases.

The results of this work, in combination with the results of Crapsi et al.~(\cite{crap}) show that the deuterium fractionation is highest at the moment of collapse. The primary use of the deuterium fractionation of N$_2$H$^+$ is to identify very young protostellar objects, because the decline of the   N$_2$D$^+$/N$_2$H$^+$ ratio is rather steep.

Comparisons of N$_2$H$^+$ fractionation with the deuterium fractionation of HCO$^+$ and NH$_3$ show clear differences, which are due either to the influence of freeze-out on dust grains or differences in chemical reactions or both. The low DCO$^+$/HCO$^+$ ratio observed by J\o rgensen et al.~(\cite{jor}) is a prediction of our model. In young, i.e. cool objects, the NH$_2$D/NH$_3$ ratio, observed by Hatchell~(\cite{nh3}) is comparable to the deuterium fractionation of N$_2$H$^+$. As the temperature increases, the N$_2$D$^+$/N$_2$H$^+$ ratio falls off much faster than the NH$_2$D/NH$_3$ ratio, which stays above 0.15 in all observed objects.  This cloud be due to desorption of deuterium enhanced ammonia from the dust grains.

\begin{acknowledgement}

We would like thank the IRAM~30m staff for their assistance during the observations. Furthermore, we gratefully thank L. Pagani for providing us the Monte-Carlo code. M.~Emprechtinger, N.~H.~Volgenau, and M.~C.~Wiedner are member of the Nachwuchsgruppe of the
Sonderforschungsbereich 494, which is funded by the Deutsche Forschungsgemeinschaft
(DFG)

\end{acknowledgement}

\end{document}